# Game Theory for Multi-Access Edge Computing: Survey, Use Cases, and Future Trends

José Moura[1], David Hutchison[2]

**Abstract.** Game Theory (GT) has been used with significant success to formulate, and either design or optimize, the operation of many representative communications and networking scenarios. The games in these scenarios involve, as usual, diverse players with conflicting goals. This paper primarily surveys the literature that has applied theoretical games to wireless networks, emphasizing use cases of upcoming Multi-Access Edge Computing (MEC). MEC is relatively new and offers cloud services at the network periphery, aiming to reduce service latency backhaul load, and enhance relevant operational aspects such as Quality of Experience or security. Our presentation of GT is focused on the major challenges imposed by MEC services over the wireless resources. The survey is divided into classical and evolutionary games. Then, our discussion proceeds to more specific aspects which have a considerable impact on the game's usefulness, namely: rational vs. evolving strategies, cooperation among players, available game information, the way the game is played (single turn, repeated), the game's model evaluation, and how the model results can be applied for both optimizing resource-constrained resources and balancing diverse trade-offs in real edge networking scenarios. Finally, we reflect on lessons learned, highlighting future trends and research directions for applying theoretical model games in upcoming MEC services, considering both network design issues and usage scenarios.

## 1   Introduction

Game Theory (GT) is a branch of applied mathematics that studies how rational players, faced with a pool of common and scarce network resources, can interact among themselves to obtain a stable allocation of system resources to fulfil the service requirements for those players. GT analyses the interaction among independent and self-interested players [1], [2].

In recent years, there has been a considerable amount of research in wireless network technologies [3]–[5], and at the time of writing one can perceive an enormous interest to apply GT in wireless networks, such as cognitive radio [6], sensor networks [7], and mobile social networks [8][9].

Data communication networks – especially wireless systems – are evolving, following a common and global trend: services and associated data, initially only available at remote Clouds, are accessible from Multi-Access Edge Computing (MEC) devices: see Fig. 1 [MEC is also called Mobile Edge Computing or Fog]. The main advantage of a system that employs MEC devices is to reduce the latency availability of both data and services to the end-users. In addition, the backhaul link overload is decreased. This occurs because most user data requests are satisfied by data previously cached at the network periphery. Simultaneously, the Internet of Things (IoT) is gradually being integrated at the network edge. In this way, new network devices are beginning to materialize (Fig. 1).

We expect, at the network edge, the typical networking operations over data (i.e. discovery, dissemination, caching) would be based on metadata, context, and environment information. By metadata, we denote QoS/QoE per traffic type or user's contract, e.g. Gold vs. Silver. By context, we mean for example device battery autonomy, data popularity, or leisure vs. business user time. By environment, we represent typically a user's social friends at the neighbourhood, device location, or mobility pattern. There are also at least two alternatives to disseminate and store the data at the network edge: reactively following the data demand, or proactively anticipating the data consumption. The previous discussion is also valid as we replace "data" by "services". Thus, we argue that GT, potentially enhanced by learning algorithms, can have an important role to model, analyse, and optimize the configuration parameters of the network algorithms or protocols deployed within each network domain. Consequently, the edge domains would operate in a more reliable way and even react appropriately to unexpected situations [10], such as load variation, conflicting strategies, incomplete information, latency, jitter, or cyber-attacks.

---

[1] José Moura is with the School of Technology and Architecture, ISCTE-Instituto Universitario de Lisboa (IUL), Instituto de Telecomunicacoes (IT)-IUL, 1649-026 Lisboa, Portugal (e-mail: jose.moura@iscte-iul.pt).

[2] David Hutchison is with the School of Computing and Communications, InfoLab21, Lancaster University, Lancaster LA1 4WA, United Kingdom (e-mail: d.hutchison@lancaster.ac.uk).

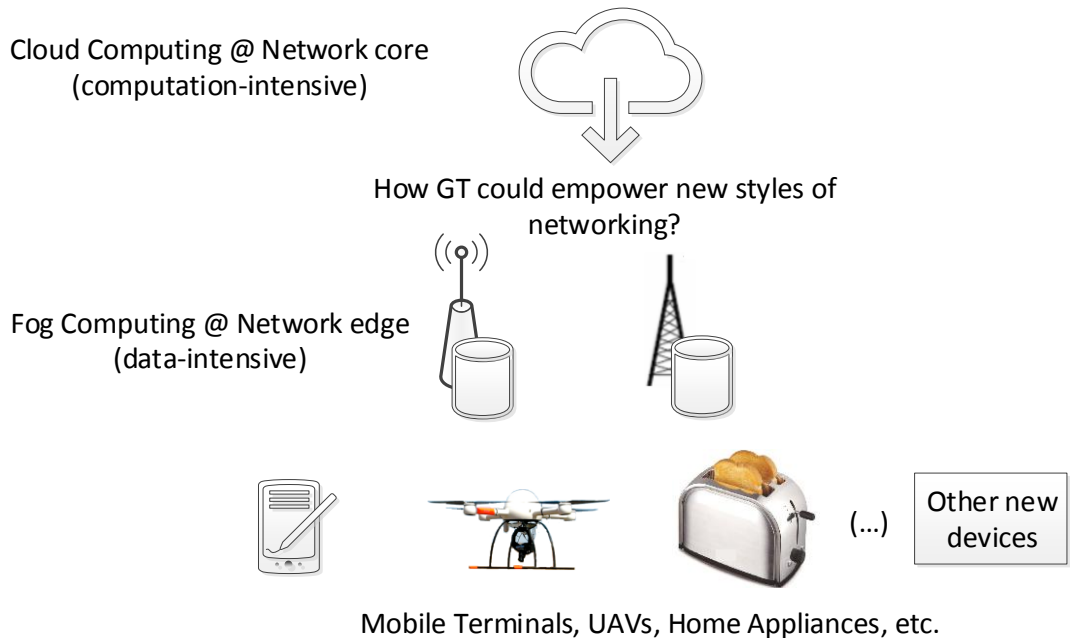

Fig. 1. MEC/Fog Emerging Environment

Our primary question (Fig. 1) is how GT can give useful hints for managing wireless data networks to fulfil the emerging challenges imposed by MEC, including energy efficiency, virtualization, storage and processing [11][12] at the network edge. We need initially to highlight the most important design aspects that have a strong impact on the supervision and orchestrated control of upcoming networks' available resources. These certainly include mobile cloud computing [11], integration of Cloud Computing and the IoT [13], the forthcoming 5G access technology [14], [15], WAVE, and SDN [16]. As an example, with the advent of IoT, mobile networks should be extended at their edge with multi-hop networks. They connect a myriad of devices, namely sensors, actuators, peer-to-peer data caches, terrestrial or aerial vehicles, and under-water or underground devices. The authors of [17] investigated a new type of Wireless Sensor Network (WSN) called a Shared Sensor Network. Further, mobile operators are planning to deploy at the edge of their networks some virtualized processing containers. These offer customers data storage and service housing with a low Round Trip Time (RTT), i.e. basically MEC [18].

Fig. 2 shows a heterogeneous wireless access network for MEC scenarios. A mobile macrocell is divided into small cells. Each small cell is like a cluster with a Cluster Header and diverse nodes (e.g. sensors). This brings some advantages, namely: i) better spectrum use; ii) enhancing service scalability by offloading traffic from macrocell to Femtocells; iii) managing data/service at the network edge; iv) diminishing the network latency/jitter and increasing the data rate; v) balancing the load among the diverse backhaul links; and vi) managing flow routing in the preferred conditions (e.g. minimum energy consumption, low delay and interference, high rate and robustness). Additional advantages include [19]: i) combining wireless and backhaul resources in the network selection; and ii) orchestrating flow admission and rate policing to guarantee end-to-end QoS/QoE. Other edge access networks visualized in Fig. 2 are discussed in subsequent sections of this paper.

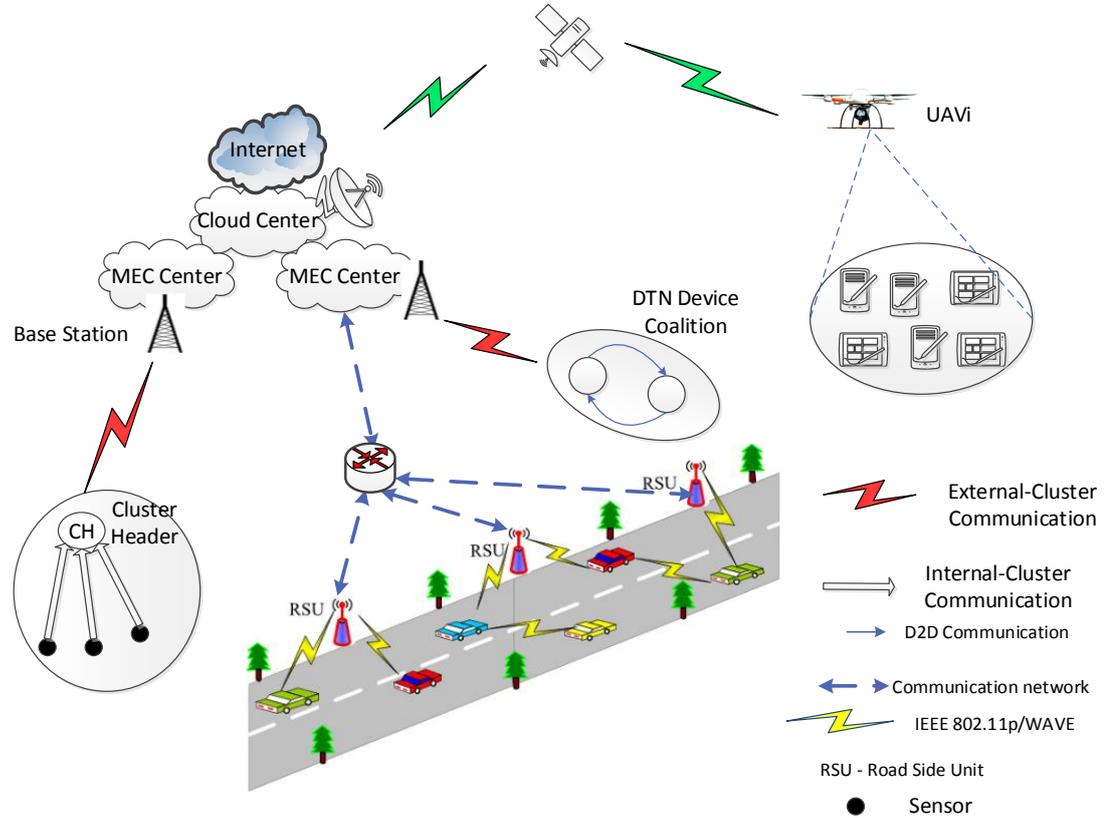

**Fig. 2. Design of a Heterogeneous Wireless Access Network involving MEC Scenarios**

To our best knowledge, the only survey about Game Theory (GT) and Multi-Access Edge Computing is [20], but it is focused on pricing models, including contract theory [21]. By contrast, our paper reviews and discusses all the relevant research directions, exploring the positive synergies of both GT and MEC in use cases of incomplete information including the aspects of learning, cooperation, and social connections. We study realistic scenarios well aligned with MEC such as heterogeneous access, small cells, device-to-device (D2D) communications, vehicular networks, IoT, energy consumption, energy harvesting, and mobile social networks. The new contributions of current work are to:

- Highlight the relevance of managing network resources to satisfy novel MEC requisites;
- Prepare a non-specialized reader with an initial background on Game Theory and MEC;
- Propose a GT taxonomy (Fig. 6);
- Compare classical and evolutionary games;
- Discuss several games namely (Iterative) Prisoner's Dilemma, Stackelberg, Evolutionary, and Bayesian;
- Debate pertinent models, e.g. cooperation enforcement and learning using classical or evolutionary models;
- Offer an interesting discussion on MEC architectures for mobile and IoT networking;
- Refresh the literature with a comprehensive review in GT, following our proposed taxonomy;
- Examine what is lacking in the literature to deploy MEC services and how GT can address the emerging requirements of MEC use cases;
- Discuss lessons learned and how model games can be applied towards MEC's expected research directions.

Table I lists the acronyms used in our survey.

**Table I Major Abbreviations**

| Abbrev. | Description | Abbrev. | Description |
|---|---|---|---|
| 4G | Fourth generation of wireless mobile telecommunications technology | M2M | Machine to machine communications |
| 5G | Upcoming generation of mobile networks | MAC | Medium access control |
| Ad hoc | Dynamic and decentralized wireless network | MDC | Multiple description coding |
| Aloha | Medium random-access techniques in both WiFi and mobile networks | MEC | Multi-access edge computing |
| BCG | Bayesian coalition game | MIMO | Multiple-input and multiple-output; it uses multiple transmit and receive antennas to exploit multipath propagation |
| BG | Bayesian game | NC | Non-cooperative |
| BOCF | Bayesian overlapping coalition formation game | NE | Nash equilibrium |
| BS | Base station | NFE | Near-far effect |
| CDMA | Code division multiple access | NFV | Network function virtualization |
| CG | Cooperative game | NTU | Non-transferrable utility |
| CN | Cognitive node | PU | Primary user |
| C-RAN | Cloud-radio access network | QoE | Quality of experience |
| CSMA/CA | Carrier sense multiple access / Collision avoidance | QoS | Quality of service |
| D2D | Device-to-device | RG | Repeated game |
| DDoS | Distributed denial of service | RL | Reinforcement learning |
| DSA | Dynamic spectrum access | RTT | Round trip time |
| DSP | Digital signal processor | SDN | Software defined networking |
| DTN | Delay tolerant network | SE | Stackelberg equilibrium |
| EG | Evolutionary game | SG | Stackelberg game |
| EGT | Evolutionary game theory | SP | Service provider |
| ETSI | European Telecommunications Standards Institute | SU | Secondary user |
| EV | Electrical vehicle | TCP | Transmission control protocol |
| FBS | Femtocell base station | TI | Tactile Internet |
| FC | Fog computing | TU | Transferrable utility |
| FD | Full duplex | UAS | Unmanned aircraft system |
| FPGA | Field programmable gate array | UAV | Unmanned aerial vehicle |
| GA | Genetic algorithm | VANET | Vehicular ad hoc network |
| GT | Game theory | VM | Virtual machine |
| GTE | Game theory explorer | WiFi | Wireless fidelity (IEEE 802.11x) |
| H2M | Human to machine communications | WLAN | Wireless local area network |
| IoT | Internet of things | WSN | Wireless sensor network |
| LA | Learning automata | UE | User equipment |

The rest of the paper is organized as follows. Section 2 presents in some depth the fundamental concepts about GT and MEC, which are required to follow the subsequent discussion. In section 3, we review the work related to the application of GT to wireless networking, and we survey MEC contributions. Section 4 discusses relevant research challenges for applying GT to emerging MEC scenarios. Finally, section 5 concludes with an analysis on the lessons learned, including a considerable number of hints for further developments combining both GT and MEC research areas. Readers already familiar with the basics of GT and MEC could now proceed directly to section 3.

## 2 Background on Game Theory and Multi-Access Edge Computing

This section provides necessary background information on GT and MEC.

### 2.1 Game Theory Overview and some Relevant Enhancements for Multi-Access Edge Computing

#### 2.1.1 Game Definition

A classical game can be represented in normal form. This assumes a tuple with three elements, (N, A, u), where:
- N is a finite set of *n* players, indexed by *i*; these players make the relevant decisions (i.e. action choice) when the game is being performed;
- $A = A_1 \times \ldots \times A_n$, where $A_i$ is a finite set of actions available to player *i*. Each vector $A_i=(a_1, \ldots, a_m)$ is designated an action profile for the player *i*;
- $u = (u_1, \ldots, u_n)$, where $u_i$ is a real-valued utility (or payoff) function for player *i*. The payoff is a kind of reward a specific player receives at the end of the game. This reward is constrained by the decisions of other players.

A very popular game is the Prisoner's Dilemma. Fig. 3 shows the normal (or strategic) form representation of this game between two players with complete information. In a complete information game, each player is aware of all the elements of the game's matrix representation. Otherwise, the game is incomplete. In Fig. 3, each player has two possible pure (static) strategies: cooperate or defect. Any payoff values respecting the next conditions, c > a > d > b, define an instance of the game. We now explain how the game is played. If player 1 chooses to cooperate and player 2 to defect then player 1 receives a benefit "b" and player 2 receives a benefit "c". In addition, each player before choosing its strategy does not have any information about the other player's decision. In these conditions, regardless of the other player's strategies, each player always selects defect and the equilibrium status of this game is {(d,d)}. Although reciprocal cooperation gives both players a better payoff, we conclude that self-interest leads to an inefficient outcome.

|  | Player2 - Cooper. | Player2 - Defect |
|---|---|---|
| Player1 - Cooper. | a, a | b, c |
| Player1 - Defect | c, b | d, d |

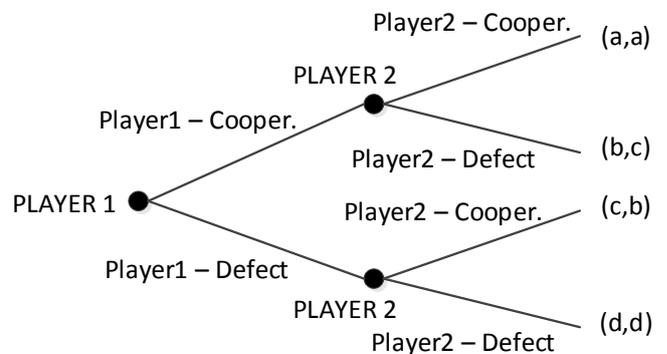

Fig. 3. Normal Form Representation of the Prisoner's Dilemma Using a Matrix of Payoffs

Fig. 4. Extensive Form Representation of the Prisoner's Dilemma

The main advantage of analysing a GT model is to discover if its players can select a set of strategies for which that model reaches a viable and stable state (in some cases also the optimum state) for the system under study. A common way to analytically discover the existence of that viable and stable state is using the definition of a pure-strategy Nash Equilibrium (NE) of a non-cooperative game. This represents no player has an incentive to unilaterally deviate from its current strategy, given that other players' strategies remain fixed [3]. Applying the previous NE definition to the zero-sum 2x2 game visualized in Fig. 5, one can verify that for any combination of strategies, each player has always an incentive to unilaterally deviate. Consequently, this game has no pure-strategy NE [3]. To solve this, we use a more generic model, where a player can select each pure strategy with a certain probability (i.e. mixed strategies). We assume player1 selects strategy *Up* with probability $p_U$ and player2 selects strategy *Left* with probability $p_L$. To evaluate the mixed-strategy NE for this game, we use the lemma described in [1][3]. Applying this, we obtain (1) and (2) below, which use the Expected Utility (EU) per strategy. Obtaining the mixed NE of this game means, as an example, player1 finding the best-response with a mixed strategy ($p_u$, $p_d$), making player2 indifferent between selecting 'left' and 'right' ($EU_L=EU_R$). Solving (1) and (2), the mixed-strategy NE is ($p_U=5/6$, $p_L=2/3$), which guarantees the payoffs of $EU_U=-1/3$ and $EU_L=1/3$, as shown in Fig. 5. Ref. [22] studies the routing over a congested network, using a mixed-strategy NE.

(1) $EU_L = EU_R \Leftrightarrow p_u * 0 + (1 - p_u) * 2$
$= p_u * 1 + (1 - p_u) * (-3)$

(2) $EU_U = EU_D \Leftrightarrow p_L * 0 + (1 - p_L) * (-1)$
$= p_L * (-2) + (1 - p_L) * 3$

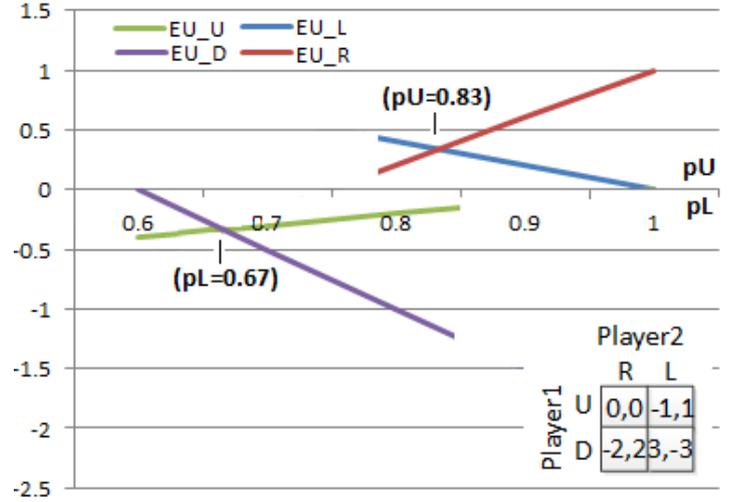

Fig. 5. Expected Utility (EU) per Player's Strategy Varying with Associated Probability ($p_U$, $p_L$) and Mixed-Strategy NE for Zero-Sum 2x2 Game

Suppose now an external observer analyses the game. This observer aims to find the strategy profile that optimizes the sum of players' utilities. This strategy is the game's Pareto optimum. As an example, one can imagine a system allocating a specific resource among the players. Pareto optimality is a state of allocation of resources in which it is impossible to make any one player better off without making at least one player worse off. Comparing both NE and Pareto designs, on one hand, NE is typically obtained using the same algorithm but instantiated among diverse players. This design normally involves a high cost to obtain the required result due to players behaving selfishly, leading to less efficient solutions. On the other hand, Pareto evaluation requires a more centralized design and assumes some cooperation among players. Due to this cooperation, the cost to obtain the aimed result is minimized. So, one can measure how inefficient the NE solution is, in comparison to the Pareto one. This is evaluated by the ratio of respective costs, i.e. the Price of Anarchy [23].

Using the normal form representation, the game typically involves a single stage. There is an alternative representation, designated as the 'extensive form', where a dynamic game has several stages, and each player chooses a strategy at each stage (Fig. 4). They are based on a tree representation that models sequential interactions among the players following a player-prioritized principle. The priority principle denotes the game designer chooses the first player to decide about the action it should perform. Then, the second player makes its choice. At the end of each tree branch, there is a terminal node with a final game result that depends on all the previous stages that were played in a sequential way, starting from the tree root and finalizing on the terminal node. The previous game is designated as 'extensive form with perfect information'. As already explained, in case a player cannot observe what the other has decided, then the players make their choices in parallel and the game is alternatively identified as 'extensive form with imperfect information'. The extensive games could be useful for modelling wireless resource scheduling in virtualized radio access networks [24]. A potential drawback of these games is they could require large amounts of memory to store the game history, essentially in scenarios with many players

7and complex interactions among them. These characteristics could hinder the scalability of extensive form games. The Game Theory Explorer (GTE) computes the equilibria of games in either extensive or strategic form [25]. There is a third game representation, coalitional form, which is discussed below.

In the following sub-section, we present our taxonomy for theoretical games and briefly explain each game type.

### 2.1.2 Taxonomy for Theoretical Games and a Brief Introduction to each Game Type

This sub-section discusses the most representative games found in the literature, organized in two distinct perspectives: classical and evolutionary (Fig. 6). The classical GT essentially requires that all the players make rational choices among a pre-defined set of static strategies. By contrast, Evolutionary GT (EGT) states that the players are not completely rational, the players have limited information about available choices and consequences, and the strategies are not static (the strategies evolve). The system's equilibrium (i.e. game solution) in both classical and evolutionary models means a stable system configuration, satisfying diverse system requirements and from which no player aims to deviate. When a player deviates from the system's equilibrium then that player is penalized on its payoff (classical game) or fitness (evolutionary game). Table II compares a classical game with the corresponding facets of an evolutionary game.

**Table II Comparing Main Characteristics of Classical and Evolutionary Games**

| Classical Game | Evolutionary Game |
|---|---|
| Players (They could be either system entities or optimization objectives with conflicting trends – e.g. delay vs. energy [26]) | Individual Organisms |
| Players intentions on rational strategies | Players intentions on heritable phenotypes |
| Player's Strategy Set | Set of all Evolutionarily Feasible Strategies |
| Payoff | Fitness |
| The Player's Strategy Set is static | Players inherit their strategies and sometimes acquire a novel strategy as a mutation (i.e. Player's Strategy Set is dynamic) |
| Focus on rationality | Focus on survivability (natural selection of the fittest group of organisms) |
| Stable equilibrium strategy optimizes either individual payoff (NE) or group benefit | Evolutionary group strategy ensures stable and high fitness (Evolutionary Stable Strategy) |

The classical GT can be divided into two dominant branches, namely NC and cooperative (Fig. 6). The essential distinction between these branches is that in NC GT the basic modelling unit is the individual player, while in cooperative GT (discussed below), the basic modelling unit is a group of several players sharing a common goal but in competition with other groups. The fundamental modelling unit of a NC game is the individual player, including her/his knowledge about the game status, expectations, and possible static strategies. The main goal of this model is to evaluate whether there exists a reasonable solution for that game. This solution implies a set of strategies that the players would rationally select for maximizing their own utility or payoff. At this point, it should be clear that a NC game is defined as a model in which any desired cooperation must be self-enforced. In addition, the players make their decisions simultaneously. So, each player normally has no information about the decisions of others (i.e. imperfect information). An exception is offered by the Stackelberg game (SG) (or Leader vs. Followers; see Fig. 7). It is a NC game type with two main differences from the other NC games, which we next explain. First, a SG game is played in sequential steps as opposed to most other NC games that are played in a single-shot way. Second, a SG has one player denoted as leader with the highest priority to take the first action. Before doing so, the leader observes the other players' (i.e. followers') strategies. Then, the leader announces its preferred strategy to the followers. The followers observe the leader's action and adapt their strategies to minimize their own cost. After this step, the followers announce their strategies again to the leader. Thus, we can define a SG as a sequential model with hierarchical decision-making that analyses the interaction between a leader (or leaders) and a set of followers to achieve a specific set of model goals (Fig. 7). The final aim of this game type is to discover the Stackelberg Equilibrium (SE), i.e. (Strategy_leader, Strategy_follower). At this point, we can conclude that SE is an evolution from a NC traditional game, where the former model adds two novel aspects to the latter one: action observation and stage repetition. Further details are given in [3].

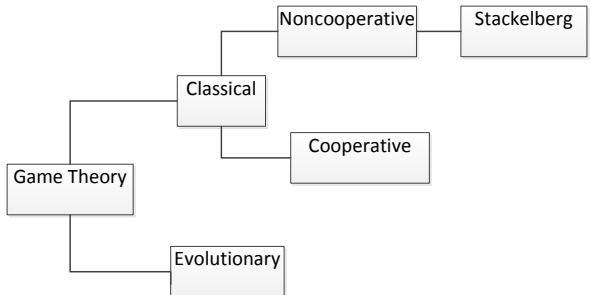

Fig. 6. Taxonomy of Model Games Used in our Work

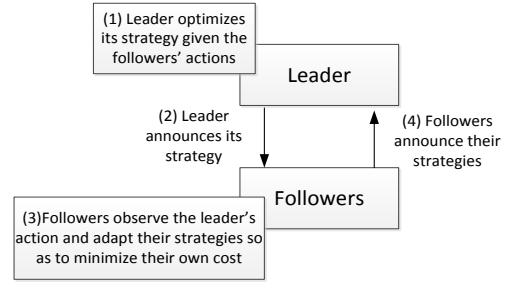

Fig. 7. Steps of a Stackelberg Game

It should be noted that a repeated game (RG) is played through several sequential stages (as we just discussed in relation to SG). A historical log is also kept, with all the players' decisions throughout the several iterations of that game. We can also classify repeated GT as an extensive form of GT, meaning a player has to take into account the impact of her/his current action on the future actions of others [27]. This special behaviour can enforce some cooperation level among the players, including in the players with no initial intention to cooperate (i.e. selfish players). The authors of [23] propose a taxonomy of RGs in wireless networks.

Cooperative Games (CGs) are suitable for designing fair, robust, practical, and efficient cooperation strategies in communication networks. The main aim behind a CG is that several players share a common objective and they can do better as a group than working alone. A key design issue in CGs is the trade-off between the stability of the coalitions and the network efficiency [3]. Notably, there are important differences in how the payoff is allocated to each player depending on whether the game is either a NC with a specific incentive to cooperation or a CG. In fact, as a NC game is used, each player receives its own payoff. In a CG the payoff of a coalition (i.e. group of players forming a single cluster) is divided among all the elements of that group following a Transferrable Utility (TU) [28], [29]. Alternatively, the payoff is not divided among the coalition members because the game has a Non-Transferrable Utility (NTU). This implies that the individual payoff of each player cannot be given or transferred arbitrarily to other players of the same cluster due to practical impairments [30][31][9]. Regardless, each player gets a benefit that depends on the actions of other players sharing the same coalition. We can consider three types of CG. The first is known as a 'canonical game'. The main goal of this game is to get the grand coalition of all users. The major potential problem is how to stabilize that grand coalition. The second type is a 'coalition formation game'. These games assume that a coalition brings some gains to its members but these are constrained by the initial cost to form that coalition. So, the key questions are how to form an adequate coalitional structure (topology) and, how to study its properties. In addition, the algorithm used by a coalition formation game typically has two functions, i.e. to merge and split coalitions. The convergence time of these functions is also critical for the game's scalability. The third and last type is a 'coalitional graph game'. In these games, the players' interactions follow a communication graph structure. The fundamental question associated with this game type is how to stabilize the graph structure. In addition, the interconnection between the players strongly affects the characteristics as well as the outcome of the game [32].

An evolutionary game is played repeatedly among some elected agents with evolving strategies. The two major mechanisms associated with the evolutionary process along the diverse population generations are mutation and selection. On one hand, the mutation (static) mechanism models the aspect of population diversity. On the other hand, the selection (dynamic) mechanism is used to promote the genetic code of agents with higher fitness than other agents. In this way, the lower fitness agents tend to disappear as the evolutionary process continues. Applying evolutionary algorithms to theoretical games allows a player with limited-rationality to select an initial strategy (action to cooperate or defect) and apply it to a specific environment (network). Then, the player receives a feedback signal from the environment (punishment or reward). If the feedback is positive then the fitness function of that player increases, meaning what that player experienced and learned (knowledge) would be transmitted to the next generation with a high probability. Alternatively, if the environment feedback is negative, the fitness function of that player decreases, implying the player's knowledge would probably not be transmitted to the next generation. After playing an evolutionary game through many generations, it is expected that most players forming the youngest generation should have a set of strategies completely adjusted to their environment. In this situation and considering that all these players belong to the same system (environment), we can consider that the evolving model has allowed the system to attain a specific state (set of objectives compatible with the youngest strategies that evolved from the seminal strategies).

We have found work that explains how evolutionary game theory (EGT) can analyse a huge variety of networking scenarios. The authors of [4] discuss applications of EGT to distinct network types such as wireless sensor, delay tolerant, peer-to-peer and wireless in general, including heterogeneous 4G networks and cloud environments. In addition, [3] studies selected applications of EGT in wireless communications and networking, including congestion control, contention-based (i.e. Aloha) protocol adaptation, power control in CDMA, routing, cooperative sensing in cognitive radio, TCP throughput adaptation, and service-provider network selection. Work about finding stable states on evolutionary games can be found in [33]. The authors of [34] discuss the stability of a RL-based distributed mechanism for strategy and payoff learning in 4G networks based on evolutionary game dynamics. The impact of evolutionary games in future wireless networks is analyzed in [35][36].

The games we discussed assume the players have complete information about other players' strategies, including in some cases the knowledge of other players' choices (e.g. Stackelberg game). But in some real scenarios, there is a lack of information about the game environment. For these cases, we should use an alternative game model, i.e. a Bayesian game (BG). In this game, the players have incomplete information about the environment they face [4]. This can occur due to some practical physical impairments that counteract the global dissemination among the nodes, e.g. channel gain information [37]. Following Harsanyi's work, a BG has a special player with random behaviour, i.e. 'Nature'. This changes the game type from Incomplete to Imperfect Information, where Imperfect Information signifies the history of the game is not available to all players. These games are called Bayesian because they require a probabilistic analysis. Players have initial beliefs about others' payoff functions. A belief is a probability distribution over the possible types for a player. Then, the initial beliefs might change based on the actions the players have taken. Comparing a BG with a non-BG (both in normal form), the latter requires only the specification of strategy spaces and payoff functions, while the former requires additional beliefs for every player [38]. In [39] a Bayesian mechanism design is explained. As a game with incomplete information is repeated, the folk theorem [40] can find its social-optimum solution. Table III illustrates the main differences among the game types in terms of whether they are static or dynamic, parallel or sequential, or complete versus incomplete.

**Table III Main Differences among the Discussed Game Types**

| Game Type | I | II | III |
|---|---|---|---|
| NC | S | P | C |
| SG | S | Seq | C |
| RG | D | Seq | C / I |
| CG | D | P | C |
| EG | D | Seq | I |
| BG | D | P | I |

Legend: I= S (static) | D (dynamic); II= P (parallel) | Seq (sequential); III= C (complete) | I (incomplete)

The next sub-section investigates how both classical and evolutionary games can be augmented to enforce high levels of cooperation among their players.

### 2.1.3 Game Theory Augmented with Cooperation Incentives

Cooperation among game players is essential for the deployment of MEC. Where a NC model is being used, the default behaviour (selfish) of NC players can be changed to be completely opposite, where the players collaborate amongst themselves. To enforce this change of behaviour, the game model should incorporate an external mechanism such as pricing [41]–[44], [45], [46] reputation [23], [47], [48], or others [21][23][49][50][51][52] [53]. These external mechanisms can enforce collaboration among diverse players to control the network operation in very distinct aspects such as either network security or the number of connected users. A pricing scheme to give a strong incentive for collaboration is discussed in [45]. The authors studied the problem of local cooperative application execution for mobile cloud computing as opposed to moving the task execution to a remote cloud. The main expectation behind the local task execution is to reduce its latency completion. To encourage mobile devices to share their unused resources, an incentive scheme is designed that benefits both the tasks' owners and mobile devices, encouraging local task execution. To model this, the authors proposed a NC Stackelberg game that permits task owner (leader) to decide the price that can be offered to mobile devices for application execution. In addition, the game allows each device (follower) to announce the amount of execution units it is aiming to offer, constrained by its battery autonomy. In [44], the authors analysed a spectrum oligopoly market with primaries and secondaries, where secondaries select a channel depending on the primary price and channel transmission rate. In [46], a NC pricing power game is proposed across both frequency and space to improve energy efficiency and throughput of coordinated multiple-point cellular transmissions.

A pricing alternative to support cooperation is reputation. Reputation ranks the level of trust of a particular node within a specific community or network domain [3]. It is a hybrid mechanism because it combines incentives and punishments. Members with a good reputation, as they positively contribute to the community, can use the network resources without any limitation, while nodes with a bad reputation, as they usually refuse to collaborate, are gradually constrained of using the network resources. A potential disadvantage of a reputation-based system is the waste of node battery and communication bandwidth to support the continuous reporting of node behavior.

The authors of [49] discuss further mechanisms besides pricing and reputation to enable positive interaction among users, namely: auctions, lotteries, bargaining games, contract theory, and market-driven solutions. The authors of [50] study how incentives are provided to the participants' social friends to stimulate cooperation, rather than directly incentivizing participants themselves. In addition, repeated games [23] [54] can enforce coordination among their rational players. As an example, considering a repeated Prisoner's Dilemma (PD) game, the usage of a discount factor (i.e. delta) and a mechanism that is triggered by a player´s defection can sustain cooperation among players. This cooperation can be enforced using "delta" within the model as the probability that the game associated with that model will proceed for another stage [26]. These results based on PD games with discount factor are pertinent for modelling and studying MEC scenarios (see 2.2). In addition, [23] discusses some interesting scenarios (also valid for MEC), related to repeated PD games with delta factor, covering: i) cellular and wireless LANs (multiple access control, mitigate security issues, manage QoS) ; ii) wireless ad hoc (packet forwarding, cooperative transmission, resource sharing in P2P networks); iii) cognitive radio (spectrum sensing, spectrum usage, spectrum trading); iv) wireless network coding; v) fiber wireless access; and vi) wireless multicast. Additionally, open research issues in discount repeated games are: i) study scenarios with myopic information; ii) payoff variations (taming defection); iii) considering a distinct delta per player; iv) trigger punishment only during a limited set of stages (i.e. forgiving after T stages following a deviating behaviour).

However, to enforce distributed coordination among the players requires the exchange of related data among them. This data is used by each player to evaluate its utility function, influencing the strategy that the player chooses within the game model. Consequently, some extra control (or signalling) messages are sent among the players, which could overload resource-constrained networks. To mitigate this issue, [53] reduces network overhead while keeping the desired control. Cooperative behaviour among players is enforced for mobile social [51] and multi-hop wireless networks [52]. In [55], the authors examine how Base Stations (BSs) can adjust their contention window to tune the listen before talk algorithm. This algorithm enables a BS, before any transmission, to sense if the unlicensed spectrum (WiFi) is free or not. When the WiFi spectrum is free, the BS transmits. Otherwise, the BS postpones the transmission (i.e. it does a backoff). The goal is to get altruistic spectrum coexistence between unlicensed and licensed bands. This coexistence is called 'license assisted access'. The authors study this coexistence using a coalition formation game. As spectrum coexistence is efficiently operationalized, the massive growth in data mobile traffic can effectively be supported.

As is well-known, the Prisoner's Dilemma (PD) is an important model to study the evolution of cooperative behavior in populations of selfish players. If the PD is played only once, it always pays to defect (selfish behavior), even though both players would benefit by cooperating. However, if the same game is repeatedly played, other strategies that reward cooperation can dominate defection. Thus, the result of an iterated (evolving) PD confirms which strategies naturally arise and remain dominant at the end of successive generations of players interacting with their own environment and opponents.

Using the results from these evolutionary models can enhance the performance of MEC scenarios (e.g. multi-hop wireless networks), by enforcing cooperation among players in a decentralized way [56], mitigating routing misbehavior induced by players' selfishness [57], and studying the evolution of trust [58]. However, the reader should be aware that as the number of players increases and iterated PD models are used, augmented by evolutionary strategies, it is harder to evolve and study cooperation for all game players [23].

We next discuss GT and automatic learning.

### 2.1.4 Game Theory and Automatic Learning

The design of wireless networks is challenging due to their highly dynamic environment that makes the discovery of stable and suitable model states a difficult task. Due to the dynamic, and often unknown, network status, modern wireless networking proposals increasingly rely on automatic learning algorithms. These learning algorithms can help to find more easily the equilibrium state of a model game.

Genetic Algorithms (GAs) are well known for their remarkable generality and flexibility when applied in a wide diversity of scenarios in wireless networks [59]. A GA models how a population of entities with similar characteristics (i.e. species), over several generations and through a process of genetic inheritance constrained by mutation, crossover and natural selection, can evolve towards a future generation of that species, formed by most of the entities fully-specialized in solving a specific problem. In applying GAs to the area of wireless networks, one can conclude that discovering the best solution for these networks could penalize their performance. This performance degradation arises due to the high GA's convergence time to find the stable solution of a specific problem, as well as the energy depleted on each mobile device's battery. For real-time applications, the use of GA to enable learning in a wireless sensor network (WSN) function (e.g. routing, or localization) is not recommended. In this context, other alternative learning techniques, such as Reinforcement Learning (RL) [60] perform better than GA, thus increasing WSN lifespan. RL is an unsupervised learning technique where each entity does not inherit any knowledge from the previous generation but learns only by itself through direct experience and interaction with the environment. Nevertheless, in some scenarios, there are strong dependencies among players (e.g. radio interference). In these cases, RL is not a realistic solution. Therefore, the initial choice (e.g. RL, GA) to implement automatic learning in each player is a critical aspect. That choice should be made according the specific scenario under study. In some cases, a hybrid solution combining RL and GA could be also viable.

In spite of the GA issues that we have just discussed, a strong advantage of GA is its easy deployment in digital signal processors (DSPs) or field programmable gate arrays (FPGAs) [59]. Aligned with this, a recent contribution uses GA to enhance the WSN lifespan, employing a new rotated crossover combined with diverse mutation operators [61]. The authors of [62] join the disciplines of machine learning and game theory to support both inter-frame coding and rate control optimization in high efficiency video coding (HEVC). A tutorial about machine learning for wireless networks with artificial intelligence is available in [63]. The authors of [64] study how self-interested agents based on some contextual status (e.g. resource abundance) independently adjust their cooperative and defection policies. These dynamic policies change the payoff values of the two possible agent actions, i.e. cooperate or defect. Each agent plays an iterated PD game within a proposed sequential social model. The results highlight some factors that, via learning, reinforce (or tame) the cooperation level among agents. A discussion about challenges and open issues in model-free learning for cognitive radio networks is available in [6]. In addition, we have found very recent games with incomplete information (i.e. Bayesian) that enforce learning among the game players in highly diverse networking use cases: hierarchical small cells [65][66][67][37][68], D2D communications [69][31][70], vehicular scenarios [30][37][71][72][73], and wireless sensors [74][75][76].

The next section discusses some of these scenarios in the context of MEC and, in doing so, introduces MEC itself.

## 2.2 Multi-Access Edge Computing Foundational Aspects

In the last few years, the huge popularity of mobile devices and the enormous increase in mobile Internet traffic have been pushing many innovations in wireless communications and networking. More precisely, the deployment of more dense wireless cells and the advent of 5G [15][10] access technology promise mobile users the eventual prospect of gigabit wireless access to cloud data with only a few milliseconds of latency. However, this remote access to data has in the present day an inherent limitation due to propagation delay values in the range of 50-200ms. These delay values are very high for latency-critical mobile applications, notably interactive ones. Recently, a new approach has emerged that aims to diminish the data access latency. This new research area is called Fog Computing (FC) [77] (proposed by Cisco) or Multi-Access Edge Computing (MEC – coined by ETSI and used preferentially in this paper) [18]. MEC is also called Mobile Edge Computing: in section 2.2.2, FC and MEC are compared. This proposal aims to integrate the concept of Cloud Computing, namely both storage and computation (virtualized) elastic aspects, into the edge of the network infrastructure, offering low latency to applications where time-efficiency is critical.

The emerging MEC approach aims to support, at the network periphery, ubiquitous and efficient access applied to a small number of system resources, such as data and/or service storage (e.g. distributed cache, proxies), computational capabilities, and data dissemination. MEC can facilitate a range of operational aspects: enhancing network coverage and capacity, load balancing at backhaul links, orchestrating the dynamic usage of available network/computing/storage resources according to the QoS/QoE of end-users, diminishing the depletion of energy at battery-operated devices, increasing network lifetime, and improving data routing based on customers social connections. MEC can also potentially provide high levels of scalability, reliability, elasticity, automation, and security [20].

The models for distinct entities of MEC systems are discussed in [18], including for computational tasks, wireless data communication channels and networks, as well as computation latency, delay on data availability, energy consumption of mobile devices, or MEC servers operating at the network edge, and privacy-aware services.

The reader can obtain further information about MEC from different perspectives, such as standardization [78], industry [79], and academia [80][81]. As already stated, our main contribution is to envisage and discuss how GT can be used to empower the gradual deployment of emerging MEC scenarios without penalizing the satisfaction of players' expectations. In addition, a valuable aspect of creating a game model is the design of utility functions to evaluate the players' payoffs associated to the available set of strategies. These utility functions should be very well aligned with the final outcomes of each game as well as the specific scenario requisites to be satisfied. A taxonomy and research challenges of utility functions for strategic radio resource management games is available in [82]. Also, [83] proposes an algorithm to pair strategic agents in a generic two-sided market without the conventional allocation of utilities among users. This solution is proposed for cognitive radios and ensures privacy among users.

The next sub-section discusses standardization work in MEC.

### 2.2.1 MEC from the Standardization Perspective

In standardization work there is interesting discussion about the more relevant MEC use cases [78][79]. ETSI proposes diverse MEC service categories [78], namely consumer-oriented, operator and third-party, network performance and QoE improvements. First, consumer-oriented incorporates services that benefit directly the end-user. Second, the operator and third-party services take advantage of computing and storage facilities at the network edge. Third, the category of network performance and QoE improvements directly assist the network service. Thus, the user-perceived quality from an application is transparently improved (such as DNS or Akamai CDNs). We next discuss MEC use cases from standardization to derive useful requirements and design constraints.

Consumer-oriented services
The MEC use cases grouped in "Consumer-Oriented Services" are listed in Table IV. We also highlight the system design constraints associated with each use case.

Table IV System Design Constraints Imposed by MEC Consumer-Oriented Services

| Use case | System design constraints |
| --- | --- |
| Gaming and low latency cloud applications | Latency, jitter, computing capacity, storage, host mobility support, transfer of user session |
| Augmented reality, assisted reality, virtual reality, cognitive assistance | Latency, jitter, dynamic virtualization, dynamic contents, user's location, environment sensorial data, response time, transfer of user session, cross-layer programming |

The first use case is related to game server applications running at the edge of the network. Here, a new kind of low latency-based game will become available to end-users. To offer this, the network device where the game server is running should have enough resources in terms of storage and computing power. While the game is running, one or more users might move around, and be connected to a different radio node (handover). As this is occurring, the connectivity between the UE and the application needs to be maintained. As the user moves away from the original location, the latency between the UE and the application is likely to increase and exceed its maximum. To avoid this, the user session might be relocated to another game server at a shorter distance from the current user location than the initial server.

Augmented reality (AR) permits end-users to have extra information from their environment by collecting sensor data, device location, and/or camera information. This information is sent to a server, located at the network edge. This server then derives the semantics of the scene, augments it with knowledge provided by databases, and feeds it back to the user device within a very short time. A recent AR contribution [84] aims for low mobile energy consumption.

Assisted reality is like augmented reality, but its purpose is to actively inform the user of any matter of interest to her/him. This might be used, for example, to support people with disabilities or the elderly to facilitate a safe and secure interaction with their environment.

Virtual reality is similar to augmented reality, but its purpose is to render the entire field of view with a virtual environment either generated or based on recorded/transmitted environments. This might for example be used to support gaming implementations or remote viewing while a user wears an input/output device (e.g. optical head-mounted device).

Cognitive assistance takes the concept of augmented reality one step further by providing personalized feedback to the user on any activity that the user might be performing. As an example, the server located at the network edge can automatically send useful information based on the user's context measured by local sensors. Thus, the data fusion, analysis of the scene, and the final system advice need to be done within a very short time. Otherwise, the cognitive assistance would be completely meaningless.

The user session needs to be personalized for all the cases mentioned above. The session continuity needs to be always maintained despite user mobility patterns. In addition, clients are not permanently using mobile networks to have access to their services. In some cases (e.g. at home or work), they might access their services located in a cloud environment through the local WiFi. However, when a user moves away from her/his indoor environment, the session needs to be exchanged without any disruption from the cloud to the MEC server co-located with the mobile cell covering the outdoor position of that user. This is a service session following up the mobile user in a ubiquitous, reliable and transparent way.

Operator and third-party services
The MEC use cases grouped in "Operator and Third-Party Services" are listed in Table V, including their constraints.

Table V System Design Constraints Imposed by MEC Operator and Third-Party Services

| Use case | System design constraints |
| --- | --- |
| Vehicle-to-infrastructure communication | Latency, vehicle mobility support, reliability |
| Unified enterprise communications | Latency, throughput, reliability, elastic service, location and other environment data |

MEC can be used to extend a connected vehicle cloud into the highly distributed mobile networking environment, caching data and services close to the vehicles that are moving within a city. This can reduce the RTT of data access which is very useful to optimize the safety and efficiency of road traffic. MEC applications can run on virtualized servers that are deployed at intelligent lamp posts along the streets, acting as WAVE APs. These applications can receive local messages directly from sensors not only within vehicles but also positioned along the roadside infrastructure, analyse them and then propagate (with extremely low delay and reliably) hazard warnings and other latency-sensitive messages to other cars in the same geographical area. This facilitates a nearby car to receive data in a matter of milliseconds, warning the human driver very rapidly (or the software agent driving the autonomous vehicle) about imminent problems.

MEC opens services to consumers and enterprise customers as well as to adjacent industries over the mobile network. The goal is to develop favourable market conditions that create sustainable business for all players in the value chain, and to support global market growth. To this end, a standardized, open environment needs to be created to allow the efficient and seamless integration of such applications across multi-vendor MEC platforms. These real-time applications should be offered with low latency, high throughput, and awareness of location and other users' environmental information.

Network performance and QoE improvements

The MEC use cases grouped under "Network performance and QoE improvements" are listed in Table VI. Also highlighted are the eventual system design constraints for each use case. At this point, GT can enhance the usage of the networking edge resources to enable these use cases.

The video management application transcodes and stores captured video streams from cameras received on the mobile cell uplink. The video analytics application running at the MEC server processes the video data to detect and notify specific configurable events e.g. object movement, lost child, abandoned luggage, etc. The application sends low bandwidth video metadata to the central operations and management server for database searches to fulfil the needs of certain applications. These applications range from public security to smart city cases, and respectively from human authorized access (e.g. with face recognition) to car park monitoring services.

**Table VI System Design Constraints Imposed by MEC Network Performance and QoE Improvements**

| Use case | System design constraints |
| --- | --- |
| Video analytics | Computational power, latency, security, |
| Mobile edge host deployment in dense-urban network environment | Latency, compute resources, storage resources, throughput, D2D Communication, traffic offloading, relaying, intelligent network selection |

The dense-urban network deployment can be useful to support services in, for example, e-Health, media and entertainment, factory, and enterprise [79]. First, the area of e-Health requires remote monitoring of health data, smarter medication, and grid access. Second, the area of media and entertainment requires on-site live event experience and collaborative gaming. Third, the factory requires an always-connected supply chain, increased level of automation, energy management, remote monitoring, and proactive maintenance. Fourth, in the enterprise area we expect seamless intra-/inter-enterprise communication, allowing the monitoring of assets distributed in larger areas, the efficient orchestration of cross value chain activities and the optimization of logistic flows. The final goal is to facilitate the creation of new value-added services. As a partial conclusion, the use case of dense-urban network requires low latency and the mitigation of wireless network congestion. To satisfy these requirements, MEC can enable D2D communication. D2D communication is a promising add-on component for cellular networks that offers several advantages: increases spectral and energy efficiencies, could decrease transmission delay, offloads traffic from the macro BS, and diminishes the load of the backhaul link. Another useful deployment is the use of relay nodes as mobile edge hosts. The MEC can manage these nodes similarly to other mobile edge hosts, allowing the system to have further options to fulfil the application requirements (notably low latency/jitter, computation, data/service storage, and of course throughput). Each mobile device could have a local agent offering intelligent management of several wireless connectivity options. These options include: i) macrocell BS; ii) femtocell BS; iii) D2D; iv) relaying transit traffic; and v) serving as a gateway for a mobile node that is out of coverage of any BS. The connectivity decisions should be made according to the mobile device´s battery status.

The next sub-section presents and compares relevant MEC architectures.

## 2.2.2 MEC Architectures for Mobile Networking and IoT Infrastructures

We analyse architectures that can support MEC services that we already debated. Initially, we present and describe some architecture options, viz: mobile cloud computing [85][86], cloudlet [87][81], MEC [88][89], FC [90][91], and heterogeneous cloud radio access network [92][93]. We start with the architecture that stores data (or services) furthest from potential consumers.

With mobile cloud computing (MCC), data is stored in remote clouds. The user perceives high latency for retrieving data from the cloud due to both high propagation delay and low data rate per user. This latency is heightened as the backhaul link gets congested. Another negative aspect is related to data (user) privacy. Some positive aspects of this architecture are that it offers both high computational power and computation offloading, which enables energy savings in mobile devices. The authors of [85] have evolved Hadoop for supporting cloud computing on Android devices. The authors of [86] discuss how network heterogeneity impacts the MCC.

The second architecture, i.e. cloudlet, assumes the data can be stored at either VMs or lighter virtual containers (e.g. dockers). These virtual entities typically run at WiFi APs. This way, users can retrieve data with lower latency than with a typical MCC transfer. In this case, the burden on the backhaul link is also diminished. Considering the usage of WiFi, the coverage is limited, which is a potential drawback of this architecture. In addition, WiFi suffers from interference and congestion problems, essentially if too many wireless devices are located within a certain geographical area. The concept of cloudlet was introduced in [87]. The authors of [81] discuss why data proximity is a relevant requirement for edge computing applications. The main advantages are that it provides highly responsive services with low delay/jitter, scalability via edge analytics that enables data aggregation and summarization, privacy-policy enforcement, and enables taming of cloud outages via nearby fall-back services.

The third architecture, MEC-based, has a similar design to the cloudlet one but works in cellular cases. In the MEC-based architecture, the resources are available from the BSs instead of the APs. The radio coverage of a cell is more efficient and scalable than for WiFi. In fact, the former has a better controlled communication channel than the latter, with much less interference and congestion. Paper [88] surveys MEC research focused on the management which joins mobile radio and computational resources. It discusses challenges and future work in this area, namely data caches, mobility management, green operation, and privacy-aware applications. Considering the computation offloading from mobile devices to edge servers [89], it enables new applications at the user device while reducing its energy consumption. However, it brings several management challenges such as selection of proper programming models, trade-off between energy consumption and execution delay, the management of multiple offloading requests, balance the load of computing resources and communication links, or VM migration.

The fourth edge architecture, FC, is formed of several vertical layers. This hierarchical design aims to support the computational demand of real-time latency-sensitive applications of largely geo-distributed IoT devices [91]. Thus, the Fog nodes form an edge cloud among IoT devices and remote clouds. The edge cloud offers to mobile devices a set of elastic services related with both computation and storage. The authors of [90] study ways to minimize the energy consumption in a FC system without jeopardizing the system characteristics of elasticity and scalability when computational and networking resources are both managed. Assuming D2D communications, FC allows (in an extreme case) the edge cloud to be temporarily disconnected from the Internet.

The fifth edge architecture, Heterogeneous Cloud-Radio Access Network (HC-RAN), aims to offer access to extra processing resources with the minimum delay and jitter, satisfying MEC requirements. To achieve these, the BSs guarantee the QoS of cloud service transmission with proper radio resource management decisions such as admission control, cell association, power control and resource allocation [92]. It suggests an architecture harmonization between cloud-radio access and Fog networks [93]. It is a very demanding task to orchestrate the management of heterogeneous network resources. Table VII lists the mobile edge network architectures discussed above.

Further coverage of mobile edge network architectures is in [94][95]. In particular, [94] surveys key MEC features. In terms of computing, they review the literature for computation offloading, cooperation between the edge and core network, and integration with 5G. They also analyse the literature for content popularity, caching policies, scheduling, and mobility management. The technologies to support MEC are discussed in [96][97]. The authors of [96] discuss virtualization technologies such as VMs and containers, SDN and NFV as well as network slicing and smarter mobile devices. In addition, they provide an analysis of the MEC orchestration considering standalone services, service mobility, joint network and service optimization. A mobile cloud specific for vehicular networks is proposed in [98]. In this case, the most critical QoS

16requirement for wireless services is latency, especially in vehicles with high mobility. They reveal a complete series of latency control mechanisms, from radio access steering to processing the caches at BSs, to target the latency requisites in mobile scenarios. The authors of [99] discuss the existing synergies of Fog and IoT, using a new design, covering the next aspects: computing, storage, networking, and control.

There are some aspects where more research is needed to optimize MEC-based architectures, viz. low latency communications for future wireless services; distributed offloading computing over heterogeneous and hierarchical environments; proactive data caching at the network edge managed by data popularity, social user connections, available node battery energy, and mobility; and low-power wireless communications and networking for IoT.

In the next section, we review GT in the context of MEC.

**Table VII Main Technical Aspects of Mobile Edge Network Architectures**

| Technical Aspects | Mobile Cloud Computing | Cloudlet | Multi-Access Edge Computing - based | Fog Computing | Heterogeneous Cloud Radio Access Network |
|---|---|---|---|---|---|
| Referenced work | [85][86] | [87][81] | [88][89] | [90][91] | [92][93] |
| Originally proposed by | [100] | [87] | ETSI | CISCO | [92] |
| Resource Access | Central | Hybrid | Hybrid | Distrib. | Distrib. |
| Distance to UE | High | Low | Low | Low | Low |
| Latency | High | Variable | Low | Low | High |
| Backhaul load | High | Low | Medium | Low | Low |
| Computational power | Ample | Ample | Limited | Medium | Medium |
| Storage capacity | Ample | Ample | Limited | Medium | Medium |
| Access | xG | WiFi | xG | Heterog | Heterog |
| Mobility support | Good | Limited | Good | Medium | Good |
| Coverage | Ample | Limited | Ample | Ample | Ample |
| Context awareness | No | Could be | Yes | Yes | Yes |
| Reliability | Low | Low | Medium | High | High |
| Hierarchy | 1 tier | 2 tiers | 2 tiers | 3+ tiers | 2+ tiers |
| Cooperation among UEs | No | No | No | Yes | Yes |
| Energy efficiency | High | Medium | Medium | Low | Low |
| Data (user) privacy | Low | Medium | Medium | High | Medium |

## 3 Review of Theoretical Model Games and Multi-Access Edge Computing

This section initially analyses the previous contributions discussing GT models for wireless networks. Then, we highlight the novel contribution of our survey when compared with other contributions that cover both aspects of GT and MEC. Next, we focus our literature review in applying specific game models, which are visualized in the taxonomy of Fig. 6, to diverse wireless networking scenarios that are well aligned with the emergence of MEC.

### 3.1 Review of Theoretical Model Games

We have found a significant number of contributions applying GT in wireless networks. A lot of this work reviews scenarios of single-hop access [101], namely: network-layered perspective [102]; multiple access [103]; random carrier sense multiple access [104]; radio resource management and admission control [105]; repeated games [23]; reputation-based network selection [47]; evolutionary games [35]; uplink resource allocation [106]; and network selection [107]. In addition, we have found related work with multi-hop wireless access such as ad hoc networks [108] and games to stimulate cooperation [52]. We have also found contributions in the area of wireless sensor networks [109][27]: pricing models [110]; energy efficiency [111]; security issues [7]; and clustering protocols [112]. In [113] a survey is presented of multiuser MIMO systems. We have also found surveys for cognitive radio networks [114][115]: cooperative games [116]; and opportunistic communications in hierarchical topologies [117]. In [34] the authors review the literature in aspects of game dynamics and cost of learning in 4G heterogeneous networks. There are some surveys for smart grids [118][119], vehicular networks [120], and D2D communications [121]. A survey about telecommunications is in [122]. The authors of [123] discuss evolutionary games for distributed resource allocation in self-organizing small cells. Discussed in [124][31][125][126][127][128][129] are several game theoretical contributions that are highly relevant for spectrum sharing, including a theoretical framework (taxonomy) to systematically understand and tackle the issue of economic viability of cooperation based on dynamic spectrum management [129]. The authors of [130] propose a constrained coalition formation game, where each UE is a player whose cost is identified as the content upload time. In this paper, the solution of the game determines the stable feasible partition for the UEs in the cell. Then, the proposed cooperative content uploading scheme guarantees lower upload delays than in the traditional cellular operation mode. In addition, the authors of [26] revise GT for existing cooperation stimulation mechanisms. Also discussed are important issues such as false judgment and node collusion. In addition, the authors argue that the root of these problems originates from the inability to evaluate accurately the behaviour of a node. This requires further investigation.

To the best of our current knowledge, the only literature covering both GT and MEC is [20], but it is focused on pricing models. The authors of [88][60] also cover aspects of GT and MEC but only briefly. We are not aware of other surveys that target the intersection between the areas of GT and MEC. The premise of our paper is that GT, as a powerful tool to understand and deal with conflicts, could help the design of adequate models and decisional algorithms to address successfully the technical challenges imposed on the network infrastructure by the wide range of new MEC applications.

For absolute beginners in GT, a recent work [131] offers code to reproduce some results from a GT network model, whereas we suggest [54][3][4][5] for readers interested in a thorough discussion on GT for wireless communications. The authors of [131] discuss a two-player strategic-form game designated as a 'near-far effect' (NFE) game, in which the two transmitters interfere with each other in the attempt to reach their own receiver. This simple scheme is the foundational model of many scenarios discussed in the publications we review: i) communications among neighboring cells; ii) hierarchical network formed by small cells within macrocells; iii) cognitive radio system joining primary and secondary users; and iv) device-to-device communications. The utility function of each terminal achieves a degree of satisfaction that depends both on the success of its transmission and on the energy spent to transmit at a specific power. In addition, due to the selfish behavior of the players, this game has the problem of NE being socially inefficient since the system distributes most of the available resources to the devices which can achieve higher throughputs. To solve this, three methods are presented to improve the NE efficiency of the discussed game: i) modify the utility function by using pricing as an externality; ii) repeat the game; and iii) put the players bargain their cooperation. Further discussion on these aspects is in [131]. We next discuss NC games.

**Non-cooperative Games**

From the available literature, we now focus on non-cooperative (NC) games. Table VIII lists the NC games surveyed in our work. At the end of the current section, we also discuss Stackelberg games because the followers of this model very often play a NC game amongst themselves.

We analyse the literature in diverse scenarios with selfish players. We have found some NC proposals related to: i) spectrum sharing for D2D scenarios such as public safety and vehicular communications [132]; ii) power control in either one-hop [133] or multi-hop [134] uplink cellular communications; iii) media access protocol for either minimizing both latency and energy consumption [26] or to reuse spectrum and minimizing interference [43]; iv) business models for IoT [135]; v) multi-rate and opportunistic routing [136]; vi) energy efficiency optimization in multi-cell massive MIMO networks with the presence of relay nodes [137]; and vii) allocating network resources according to expected load [138].

A very good example of a NC game where their players are not directly associated with network entities is in [26]. The players of this game are related to performance network metrics, such as energy and delay. These metrics have conflicting operational trends. In fact, the delay is minimized at the cost of increasing energy consumption. However, the energy consumption is minimized at the cost of increasing delay. Consequently, these two metrics have been formalized as the players of a NC game to solve a conflicting multi-objective optimization problem in energy-constrained, delay-sensitive wireless sensor networks. As a final game result, given the two performance requirements (i.e., maximum latency tolerated by the application and initial energy budget of nodes), the proposed model allows settings of MAC parameters (e.g. sensor wakeup period) to reach a fair equilibrium that minimizes both latency and energy consumption.

**Table VIII A Summary of Non-cooperative Approaches for the Resource Management in Wireless Networks**

| Ref. | Area | Goals |
|---|---|---|
| [105][139] [140] [141] [132] | Network resource sharing | Enhancing the usage of a common pool of constrained resources among a set of consumers, ideally in a fair way. Elastic and scalable management of resources |
| [142][137][143][133][144][65][145][146][147][148][134][149][150] | Power control | Limiting interference and increasing the Signal to Noise Ratio. Energy efficiency |
| [26][151] [152] [153] [154][155] | Medium access control (MAC) | Schedule the access to a single communications channel shared among several users. Avoid collisions. Energy efficiency. Support QoS/QoE |
| [41][43][44][156][157][135] | Business model | External incentive to enhance cooperation. Increase profit. Enhance resource use |
| [136] | Multirate opportunistic routing | Steering traffic through a network for maximizing end-to-end throughput |
| [137][143][145][146][134][158][159][160][66][161][162][26][163][164][165][166] | Energy efficiency | Trade-off between energy consumption and fulfilling QoS/QoE requisites; enhance the autonomy of battery-operated nodes; increase lifetime of a wireless mesh network; low-power wireless communications and networking for IoT |
| [167] | Security | Study about data injection attacks on smart grids |
| [138] | Resource management | Proactive resource reservations for the expected number of users |

It is well known that in a NC game the players – due to the lack of information about the intention of others – assume selfish decisions, decreasing the global network performance. In the following text, we discuss two extra mechanisms that are added to the NC game model to enhance collaboration among players: Reinforcement Learning (RL) [134], and payment scheme [136]. The authors of [134] use NC game theory paired with RL methods to investigate the problem of power allocation in the uplink communication at both source and relay nodes on a multiple-access multiple relay scenario. This model achieves a good compromise between energy efficiency and overall data rate. By contrast, [136] explores the combination of two possible features of wireless networks: multi-rate and opportunistic routing. The multi-rate feature is related to the multiple transmission bit rates specified by IEEE 802.11 protocols. The opportunistic routing allows any node that listens to a prior packet transmission to participate actively in packet forwarding. Aggregating these two features and by following the routing and incentive protocol, each node maximizes its payoff. Specifically, the incentive protocol works as follows. Added to the network are probe messages, which measure the link loss probabilities, along with a cryptographic component to prevent the probe message from being forged, and a payment scheme to guarantee that the nodes cannot benefit from manipulating the link loss probability measuring process or deviating from the routing decision. So, the spirit behind this incentive protocol is maximizing the global network performance, while accepting some network overhead due to the extra probing traffic among the nodes. Further work is required to protect it from collusion.

Another way to mitigate the selfishness of NC players is enabling the players to learn about more suitable decisions. This learning is enforced via repeating the game through several stages, i.e. NC RGs. We have identified some recent contributions in the scope of NC RGs: radio spectrum sharing among Primary and Secondary users [44]; multimedia dissemination through either a content distribution network [168] or a mobile multi-flow multicast approach [169]; mobile D2D communication [69][170]; and scheduling of beacons for competing drones and enforcing energy savings [159]. The authors of [159] study the scheduling of beacons based on a NC game for two competing drones that cover (e.g. using WiFi) two small cells with distinct device's density, as shown in Fig. 8. The backhaul communication in each drone is provided through a satellite C-band link. Each drone has two possible strategies; it is either in "idle mode", or sending beacons for mobile users on ground. The UAV's payoff is the difference between the positive outcome of the successful first contact with mobile users on the ground and the energy cost to achieve that. They study network configurations for maximizing the likelihood of getting in contact with the mobile users on the ground with the minimum energy consumption, which is vital for drones to fly as long as possible. This work has some aspects that can be enhanced: the UAV backhaul is a satellite link with high latency and limited rate; not discussed is the energy consumption in each UAV associated with the satellite interface. From all the NC games previously surveyed in the current publication, we highlight the following because they are associated with emerging usage scenarios: WSNs [26], 5G [43][132][137][145][162][163][166], UAVs [159][165], and power-grid [167]. Some NC games for wireless and communication networks are covered in [3], [4].

NC games have a relevant drawback in terms of the absence of players' learning; this can impair the discovery of a game (stable) solution, thus diminishing the game's efficiency. Stackelberg Games (SGs) try to enhance the efficiency of game results by playing the game in a sequential and hierarchized decision-making way. In this game, also implicit is a set of players with high priority that are the first ones to decide (leaders). Then the remaining nodes (followers), depending on the actions selected by the leaders, should choose their most convenient actions. The surveyed papers rely on SGs modelling some MEC use cases: i) offloading mobile computation [45]; ii) efficient cellular communications [171][172][54][173]; iii) resource allocation in hierarchical cellular networks [65][37][66]; iv) multimedia services [48][174]; and v) management of micro-grids [175] [176].

The authors of [149] study energy charging in a power system composed of an aggregator and multiple electrical vehicles (EVs) in the presence of demand uncertainty. They propose a SG, where the aggregator is the leader and the EVs are followers. They propose two different approaches under demand uncertainty: a NC and a cooperative design. Both are robust games [204]. In the robust NC case, they present the energy charging problem as a competitive game among self-interested EVs, where each EV chooses its demand strategy for maximizing its benefit. In the robust cooperative model, an optimal distributed energy scheduling algorithm is formulated; this maximizes the sum benefit of the connected EVs. They theoretically prove the existence and uniqueness of a robust Stackelberg equilibrium for the two approaches and develop distributed algorithms to converge to the global optimal solution independent of the demand uncertainty.

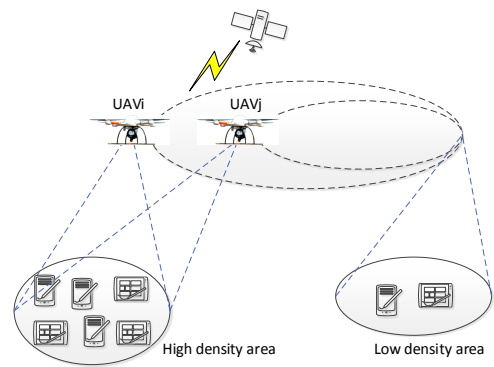

**Fig. 8. Drone Small cells**

**Summary of Non-cooperative Games**

Before choosing a NC game to support the correct operation of wireless networks, a system designer should be aware that this choice offers advantages – but it could also imply some disadvantages. The main advantages are that wireless nodes compete for a limited set of network resources; they obtain a more robust distributed control algorithm in comparison with the centralized game option; and they diminish the signalling network traffic due to local processing in each node. However, the possible disadvantages that the designer should be aware of are the absence of learning as the users simultaneously play the game; the self-interested user behaviour is naturally induced due to the lack of information about other players' actions; and a difficulty in attaining global system optimization because of the distributed system design. We have seen two ways to mitigate the problem of selfishness among players of a NC game – by incentivizing cooperation [136], or by learning [134] [44] [168] [169] [69][170] [159]. These augmented NC games can enable low-complexity distributed control algorithms for MEC dynamic use cases that satisfy harsh trade-offs, such as that between energy efficiency and overall network performance.

We have also studied SGs. Some applications of these games are: Software Defined Networking (SDN) scenarios, where the SDN controller is the Leader; Femtocell power control [3] in hierarchized mobile networks; and D2D communication [171]. The main advantage of using a SG model is to optimize diverse virtualized resources (e.g. computation, storage, and networking) of complex topologies at the network edge under users' Quality of Experience. The main problematic issues the network designer should be aware of are: implementing a robust mechanism to ensure the correct and synchronous shift among leaders and followers; Stackelberg Equilibrium (SE) could give a worst result than NE due to the hierarchical decision-making process among leaders and followers [3]; and, it requires complete and perfect information about other strategies and payoffs. In this situation, communication jitter among a leader and followers could disrupt the right control sequence and create instabilities on the control loop, affecting the model results. The next section reviews cooperative games.

**Cooperative Games**

Cooperative Games (CGs) have been applied in wireless networks to design fair, reliable, and efficient cooperation algorithms. One should be aware that to incentivize cooperation among the players of a game, it is not necessary to use a CG (see the discussion in 2.1.3 about Reputation and other incentives for cooperation). In the following sub-section, we discuss a variety of recent contributions of CGs to diverse facets of wireless access networks, viz: small cell networks [124][177][178][179][180], followed by D2D communications [31][181][130][9][182][183][184][8][185][186][187], vehicular networks [71][72][188][189], wireless sensor networks [190][191][192], cellular communications [181][193][194], and cooperative repeated games [195][196][190][30]. These are highly relevant for MEC/FC.

Small cell networks

We start with small cell networks. Small cells are used to mitigate congestion problems in 3G systems (and beyond). This choice is based on the usage of Femtocells sharing the same spectrum of the macrocells, which raises new challenges, such as the interference mitigation [124][179], spectrum management [178], uplink user association [177], traffic offloading [124], and management of the network throughput considering not only network status but also the social ties among users, essentially when D2D communications are available [180].

Considering the hierarchical system formed by macro and small cells, the interference mitigation becomes more complex to manage because it has two components [124]: cross-tier and co-tier interference. The cross-tier interference is from the macrocell BS to the small cell BSs. The co-tier interference is among small cell BSs. The authors of [124] propose a CG for the mitigation of both co-tier and cross-tier interferences as well as the resource allocation in dense heterogeneous networks. In case of congestion this allows the traffic offloading from macrocell base station to small cell base stations. Alternatively, [179] focuses specifically on the mitigation of co-tier interference among small cell BSs.

In [178], the authors enhance the performance of cognitive Femtocell networks through the hybrid overlay/underlay model, with both the non-utilized and under-utilized spectra. In addition, based on prior work, they explore both spatial and frequency reuse. They formulate the sub-channel allocation problem as a coalition formation game among Femtocell users under the hybrid access scheme with negative externalities to effectively characterize and tame the interference. Their proposal offers a core solution [197] with stable and efficient allocation.

The authors in [177] suggest a solution based on an analogy of uplink user association as an admission game between mobile users and a hierarchical cellular infrastructure formed by both small cell and macrocell base stations. They combine matching theory [198] and CG [199] to solve some potential conflicts in this scenario, e.g. giving strong incentives to the small cells to extend the macrocell coverage while maintaining the users' QoS. User mobility and multi-homing aspects are out of the scope of [177].

Device-to-device communications

Only recently, GT and CGs have also been successfully applied to address the new challenges of D2D communications in mobile scenarios (i.e. these challenges also involve 5G, IoT and mobile edge computing use cases). These new challenges are grouped into the categories of resource allocation [9][31][130][181][183][185], energy-aware [182][184][186], and social-aware [8][187].

We now detail the resource allocation proposals for D2D. In [9] the authors study a coalition formation game with Non-Transferable Utility (NTU), and the mobiles are partitioned into many coalitions, each of which applies a cooperative scheme to maximize their profit. The authors of [31] introduce a new Bayesian non-transferable overlapping coalition formation (BOCF) game to study spectrum sharing by D2D communications in cellular networks. This work was extended in [181] to the case in which a mobile device belongs simultaneously to multiple coalitions. In [130], a constrained coalition formation model considers the content uploading time and also includes the coverage constraints for the D2D links so that only feasible coalitions are formed. Article [183] investigates the scenario of uplink radio resource allocation when multiple D2D pairs and cellular users share in an optimized way the available resources, considering a cross-tier interference among primary and secondary users. The authors of [185] design and evaluate a dynamic distributed resource sharing scheme that jointly considers the mode selection, resource allocation, and power control in a unified framework, with the goal of maximizing the available rate under a series of practical constraints in a mobile cellular network that has D2D pairs and cellular users.

Next, we look at the proposals for D2D which are mainly concerned with energy-efficiency. In [182] the authors propose a CG in a scenario of wireless content distribution as an effort to minimize the overall energy consumed whenever several mobile terminals seek to download common content of interest. In [184] a simple CG is studied for energy-efficient D2D communications in public safety networks. In addition, the authors of [186] are concerned with diminishing the energy consumption on the cellular network with D2D communications.

Further, we cover D2D proposals with a focus on social-aware scenarios. Article [8] develops a coalition game-theoretic framework to devise social-tie-based cooperation strategies for D2D communications, which can achieve significant performance gain over the case without D2D cooperation. In [187] the authors propose a trust-based and social-aware coalition formation game for multi-hop data uploading in 5G systems.

A survey of interference management for D2D communication and its challenges in 5G networks is available in [200]. Article [9] reviews game-theoretic resource allocation methods for D2D.

Vehicular networks

The main aspect of a vehicular network is the great difficulty (due to vehicle mobility and limitations on the network coverage) in always maintaining high-quality coverage via a convenient access technology, e.g. WAVE (IEEE 802.11p) as visualized in Fig. 9. Consequently, a way to mitigate the previous limitation is to enable multi-hop communications in this emerging scenario. To support multi-hop communications in an efficient way, cooperation among the vehicles needs to be improved. Coalitional formation games can thus be very useful for attaining those goals. Normally the players of these games are the vehicles.

The authors of [72] propose a cooperative Bayesian Coalition Game (BCG) as-a-service for content distribution amongst vehicles. This work is complemented by [71] which proposes a NC Bayesian CG. These two contributions [72][71] could have convergence time issues due to a exponentially increasing complexity in the algorithm to form the coalitions of vehicles when the number of vehicles is relatively high. Alternatively, in [188] the algorithm to form coalitions only requires a single iteration, making it a scalable solution in terms of the number of vehicles, each of which is searching for the best access network. A limitation of this proposal is that at any given time, a vehicle can use only a single network interface. The authors of [189] investigate a reliable message delivery in Vehicular Ad Hoc Networks (VANETs). They model the cooperative service-based message sharing problem in the VANET as a coalition formation game among nodes. Some nodes within a coalition operate as relays. So, their solution, for each case, chooses the most appropriate relay to enhance rate and reduce delay. The authors of [194] develop a distributed coalition formation algorithm to create an adequate coalition structure for vehicular networks.

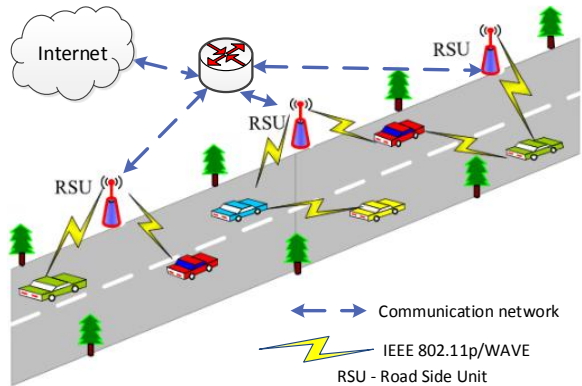
**Fig. 9. Vehicular Communications**

Wireless sensor networks
We have found several CGs to enhance the operation of wireless sensor networking environments for the following aspects: network lifetime [111], security [191][192], and MAC access [190]. The authors of [111] review GT for wireless sensor networks. The focus is on discussing games that enforce the cooperation among players for maximizing network lifetime.

In [191] a CG with a reinforcement-based learning algorithm is proposed for WSNs. It is a three-player strategy game consisting of sink nodes, a BS, and an attacker. The proposed model implements a cooperative security game to mitigate DDoS attacks. Article [192] suggests a security strategy to predict the attacks and their effect using cooperating camera sensors. The model is based on a threshold for the probability of error of the captured scene. This approach provides a solution for false alarms, attack prediction, and selfish behaviour. This work can be extended to tame the issues of harsh environmental conditions, and wrong calibrations or component defects. Nevertheless, RL in large networks could converge slowly to the stable-point operation due to high complex problem of estimating all possible actions and states of players. In these cases, assuming players with bounded-rationality could make the problem analysis more tractable.

The authors of [190] propose a cooperative GT with incomplete information to enhance the performance of MAC protocols in wireless mesh networks (WMNs). In this game, first, each node estimates the current game state (e.g., the number of competing nodes). Second, the node adjusts its equilibrium strategy by tuning its local contention parameters (e.g., the minimum contention window) to the estimated game state. Finally, the game is repeated several times to reach the optimal performance. To use the game effectively in WMNs, the authors present a hybrid CSMA/CA protocol by integrating a proposed virtual CSMA/CA and the standard CSMA/CA protocol. When a node has no packet to send, it contends for the channel in virtual CSMA/CA mode. This way, the node can estimate the game state and obtain the optimal strategy. When a node has packets to send, it contends for the channel in standard CSMA/CA mode with the optimal strategy obtained in virtual CSMA/CA mode, switching smoothly from virtual to standard CSMA/CA mode. The solution's simulation method shows promising results in terms of throughput, decrease delay, jitter, and packet loss rate. We have also found CGs addressing energy efficiency [111] and clustering [112] for wireless sensor networks.

Cellular communications
Cooperative packet transmission can enhance the throughput in wireless networking infrastructures by taking advantage of the mobility of the users' devices, essentially at networks with intermittent connectivity, high delay and error rates such the typical case of delay-tolerant networks (DTNs). Here, the DTNs (with D2D communications) are directly related to a decentralized mobile social network [9], where the data (or context information such as location) transfer can be performed locally among users or devices by means of their own mobility patterns that empowers the formation of opportunistic coalitions among the local devices, as visualized in Fig. 10. As an example, node 1.1 forms a coalition after moving to a location nearby node 1.2. Within that coalition, the two nodes can perform D2D communications.

The authors of [181] investigate a cooperative game to study spectrum sharing for D2D communications in cellular networks. This work offers the novelty of a mobile device belonging simultaneously to multiple coalitions. It is assumed that all players in a coalition abide by the entire coalition decision, but this does not hold in real scenarios if we have the same mobile device belonging to two coalitions but owned by distinct operators with incompatible management policies.

In [193] a CG theory is proposed for determining the rate region of multiple description coding (MDC). MDC introduces enough diversity at source coding level to mitigate packet losses by adding description information. Each description provides a coarse reconstruction of source information at destination. As the number of received descriptions increases, the reconstruction quality would be improved. Moreover, the Shapley value [197] is used to determine each description's rate. Thus, a fair rate adjustment is obtained that comprises entropy and mutual information of the original and reconstructed descriptions. They do not support multi-hop cellular communications.

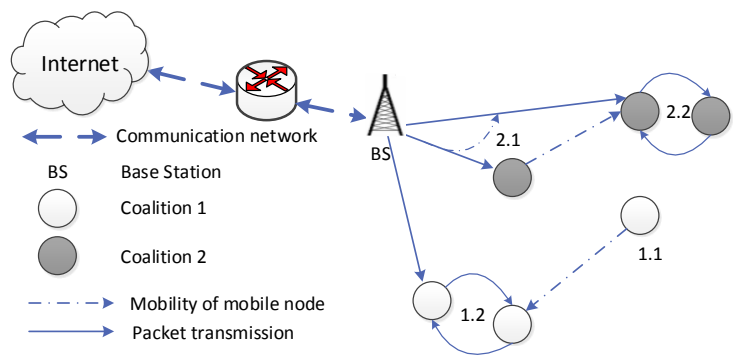

**Fig. 10. D2D Opportunistic Communications**

Cooperative Repeated Games

A repeated game (RG) [23][7] is an effective tool to avoid potential conflicts among wireless nodes. The occurrence of these conflicts is because of selfish behavior. The selfishness results in poor network performance and detrimental individual payoffs due to e.g. the interference among wireless nodes that are competing amongst themselves. An important advantage of using a RG is to enforce learning at cognitive nodes in a game with incomplete and imperfect information [201]. Thus, the game evolves to a state where the interference is tamed and QoS requirements are fulfilled.

From the previous work in the scope of cooperative RGs, we have selected some recent contributions with the following goals: i) trade-off between load balancing and energy efficiency with extra mechanism to both discourage selfishness and ensure fairness [195]; ii) increase the total throughput of a mesh wireless network because selfish nodes are encouraged to cooperate under the threat of punishment [196]; and iii) delay tolerant networks [30].

**Summary of Cooperative Games**

This sub-section has reviewed the literature of CGs at the network edge. The topics covered were: small cell networks, D2D communications, vehicular networks, wireless sensor networks, and cellular communications.

We can outline additional interesting application examples for CGs. As an example, coalitional graph games can be used for wireless topologies using either relays or device-to-device communications in two distinct phases: i) network formation; and ii) flow steering through a hierarchized network. Other possible interesting scenarios to apply CGs are: trust management in wireless networks, multi-hop cognitive radio, relay selection, intrusion detection, P2P data transfer, multi-hop relaying, and packet forwarding in sensor networks [3]. Further scenarios are: processing and radio resource allocation for computation offloading from mobile devices to edge servers; and named data networking edge systems.

The main advantage of using this type of game is that an initially large system can be divided into diverse smaller coalitions, which can simplify the global system operation a great deal. The main issue of difficulty the network designer should be aware of is that, as the network dynamic is high (e.g. due to device mobility), the coalition formation algorithm is continuously changing the clusters and the system does not stabilize in a stable coalition formation. This can lead to excessively high values of convergence time and energy depletion on battery-operated devices.

The next section reviews various evolutionary games in highly diverse networking scenarios.

**Evolutionary Games**

Applying evolutionary algorithms to theoretical games allows players with limited rationality to learn from the environment and take individual decisions to attain each game's equilibrium in a distributed way. We have found in the literature games that use evolutionary algorithms in highly diverse networking scenarios. From these, we have selected a few with the following goals: i) develop strategies to efficiently use the available network resources [202][203][204][35]; ii) distributed management of hierarchical mobile networks with small cells for enhancing the usage of radio resources [123][205][121][206]; and iii) in VANET scenarios [207]. The evolutionary games (EGs) we have selected are now discussed. We start first with EGs that enhance the usage of network available resources strategies for either wired [202] or wireless [203][204][35] cases. In [202] the authors use EGs to study the necessary network conditions to concurrent transport protocols may either co-evolve or not for dealing with the congestion problem. In the latter case, a dominant protocol would be operating in the network. In addition, the authors propose some directions in upgrading the protocols to enable the network operation at an optimum stable state. An evolution could be how to deploy, in a non-disruptive way, the protocol upgrades in a real networking system. In [203] the authors discuss an EG for vehicle-to-vehicle (V2V) caching. In V2V caching, only users who actively participate can obtain cached contents from others, and users who participate in V2V caching must keep the contents in storage after watching, using and consuming them. Thus, an evolutionary stable strategy is a mixed strategy for the whole population (i.e. the winner strategy), such that the turbulence of a small proportion of other strategies will gradually disappear in the long-term trend. Examples of these minority strategies are related to a reduced number of users leaving or joining the system. The authors of [204] consider NC mobiles, each faced with the problem of which subset of WLAN access points (APs) to connect (and multi-home) to, and how to split the traffic among them. Considering the many users' regimes, they obtain a potential game model and study its equilibrium. The authors analyze the user performance of UDP/TCP throughput with varying frame lengths over WLAN. They also obtain pricing for which the total throughput is maximized at equilibrium and study the convergence to equilibrium under various evolutionary dynamics. Article [35] reviews evolutionary CG theory applied to wireless networking and communications use cases such as opportunistic and cognitive radio networks. The authors have assumed a homogeneous network.

Second, we refer to EGs that provide distributed management of hierarchical mobile networks with small cells for either optimizing the usage of radio resources [123][205] or taming interference [121][206]. In [123] the authors study the spectral coexistence between a macrocell and Femtocells, using tools from evolutionary GT and RL. In addition, a variant of the evolutionary game approach (referred to as replication by imitation) is investigated where Femtocells probabilistically review their strategies and imitate other Femtocells in the network. They finally conclude that the spectral efficiency and convergence to a system stable state are shown to be driven by the type of information available at Femtocells. The performance of the evolutionary-game-based algorithms under information delay is highly dependent on the network parameters such as channel gains and number of devices. Other similar work that optimizes the usage of radio resources is available in [205]. In [121] two mechanisms for interference mitigation are studied, respectively EG and Reinforcement Learning (RL) and their results are compared. The results reveal that through constrained learning, Femtocells can self-organize using only local information and mitigate the interference towards the macrocell network. Nevertheless, this process could be slow to converge because learning implies a lot of "trial-and-error" steps. On the other hand, an EG can ensure a more expedited alternative to reach the optimum network configuration but at the expense of high signalling overhead in the network. In practical terms, this last alternative can be also difficult to deploy because the instantaneous information exchange among Femtocells is normally hard to achieve. Other similar work that deals with interference is available in [206]. In this context, the effect of information exchange delay on the convergence performance of the distributed resource allocation algorithm is still an open issue.

Third, we discuss an evolutionary game for VANETs [207]. The authors propose a cluster-based Public Goods Game group interaction model to investigate how the cooperation level among vehicles depends on the clustering topology. Their results show that a topology with a high number of clusters increases the connectivity among vehicles, enhancing cooperation. They also assume a minimum number of nodes for the correct operation of their proposal.

**Summary of Evolutionary Games**
We have reviewed EGs for network edge services: efficient usage of network resources, small cells, distributed learning to tame the interference issue, and enhancing cooperation in VANETs. Further interesting application examples for evolutionary games (EGs) are: congestion control; medium access control; routing-potential game; cooperative sensing in cognitive radio; TCP throughput adaptation over wireless networks (normally via the last network hop); user churn behaviour; network selection with dynamic bandwidth allocation [3]; and elastically allocating resources (storage, processing, sensor-based) from edge cloud providers to mash-up MEC applications.

The main advantages of using this type of game are that decision strategies are adapted to the environment sometimes in a non-rational way due to lack of information, and they could enable a dynamic and reinforced learning using a "trial-and-error" until either the problem is solved or a maximum number of iterations is reached. However, the main issues to be aware of are high convergence time, and the system's evolution could exterminate the "wrong species", giving unexpected solutions.

In the next section, we discuss how GT can bring positive outcomes to MEC.

## 3.2   Multi-Access Edge Computing

We continue our literature review on Multi-Access Edge Computing (MEC). We have found some contributions on MEC [208][95] covering the topics of: the communication perspective [88]; computation offloading [89]; convergence of computing, caching and communications [94]; emerging 5G network edge cloud architecture and orchestration [96]; software-defined networking [76]; architecture harmonization between cloud radio access networks and fog networks [93]; Internet of Things [99]; and latency control in software-defined mobile-edge vehicular networking [98].

Whereas none of the previous work analyzes GT applied to MEC, we discuss GT research topics related to MEC, namely on wireless sensor networks, cognitive small cells, vehicular networks, and unmanned vehicles. We also identify specific scenarios where GT can offer a positive impact on the performance of the involved wireless networks.

Wireless sensor networks
Wireless sensor networks (WSNs) have profound significance for environmental surveillance and remote monitoring by placing sensors in places difficult to access by humans. Consequently, energy-efficient algorithms and protocols need to be developed for enhancing sensor lifetime, e.g. energy harvesting [209][210]. Additionally, the sensor data demands low latency because the data signalize faulty situations in the monitored environment that require urgent action [211][212]. In [213] localization scheme named Opportunistic Localization by Topology Control is proposed, specifically for sparse Underwater Sensor Networks (UWSNs). In [7] GT is used to control security threats in WSNs. Articles [61] and [62] study economic and pricing models associated with WSNs. Papers [17][214] outline future research trends in WSNs.

We discuss below two WSN scenarios where GT can bring positive outcomes. In the first scenario, the sensors are typically resource-constrained devices in terms of both computation and their battery autonomy. In addition, these devices are deployed at remote locations, making their individual maintenance either very difficult or impossible to do. Thus, the depletion of the sensor battery needs to be minimized, enhancing the sensor's lifetime. Such a goal can be achieved by a cooperative game, where groups of nearby devices form coalitions to achieve a specific goal, and then share the reward. Moreover, by adding Reinforcement Learning (RL) [60] to the previous game, the uncertainty and lack of prior information are addressed. Therefore, the sensors can learn about the more convenient cluster-based topology that enhances their battery autonomy. In the second scenario, Blockchain technology [215] is used for digital money transactions, authenticating the source of each message. The Blockchain can authenticate the sensors before they send their messages to the network. This is relevant for protecting the sensor environment against any initial preparation of DDoS attacks. A possible way to deploy this distributed authentication framework is where resource-constrained sensors assume the role of miners. However, as the Blockchain requires heavy processing for its operation then, alternatively, that heavy processing should be offloaded from the sensors to the MEC servers. The computation offloading can be modelled using a Stackelberg game. The leader is the sensor requiring the Blockchain authentication. The followers are the MEC servers running as miners to solve the Blockchain puzzle [215].

Cognitive small cells
The deployment of low-cost and high-capacity cognitive small cells over existing cellular networks can offload traffic from the macrocell to the small cells. To achieve this, some problems induced by this hierarchical design need to be successfully addressed, such as: the share of spectrum among macrocell base stations (BSs), macrocell users, Femtocell BSs and Femtocell users [216][217][218][219][220]; energy consumption [221][222][223][224]; control of power transmission and taming of interference [225][226][227][228]; security [229][230]; enhanced cooperation among players via pricing [231]. Article [6] reviews applications of Model-Free strategy learning in cognitive wireless networks. In [232] the authors discuss relevant trends and challenges in the deployment of millimetre wave, massive MIMO, and small cells towards 5G networks. Thus, the capacity of mobile networks should increase to support the exponential growth in data traffic with minimum latency and jitter. Several deployment strategies of mobile networks can increase their capacity [233], namely: i) use of larger bandwidth by exploiting higher spectrum frequencies [163]; ii) increasing spectrum efficiency by using MIMO [161]; and iii) spatial reuse of spectrum by deploying D2D [31], small cells [218], and heterogeneous networks [66], including their energy-efficiency [234]. Open issues for D2D communications include the prevention of Denial-of-Service (DoS) attacks, particularly in D2D LANs [9]. In addition, a novel distributed method to validate user identity (e.g. Blockchain [215]) is vital to enhance D2D security. Machine-to-Machine communications is also a promising technology for future networks [235]. This allows automatic interaction among devices, namely sensors and actuators.

Next, we discuss two cognitive small cell scenarios where GT can bring positive outcomes. In the first, the data caching at MEC nodes as near as possible to their final mobile consumers minimizes the user data access latency, backhaul load, and energy consumption on the mobile devices' batteries. Such results can be obtained by a CG augmented by RL, which is a convenient solution to deploy a scalable distributed system with uncertain information. In the second scenario, each mobile device can establish distinct types of communications, such as: i) macrocell BS; ii) Femtocell BS; iii) the mobile device behaves as a relay, retransmitting messages to other devices; iv) the device behaves as a gateway node, retransmitting to a BS (macrocell or Femtocell); v) usage of either a single access technology or multiple ones (e.g. xG, WiFi, Bluetooth). We envision an agent running in each mobile device. This agent manages all these connectivity options to forward opportunistically the traffic from neighbouring devices. The agent is the leader of a SG and the followers are the distinct interfaces of the mobile device. The behaviour of these interfaces can be modelled as a NC game augmented with RL, for enhancing resource usage and traffic delay.

Vehicular networks
Interest in intelligent transportation systems and vehicular ad hoc networks has increased a great deal in recent years. The number of applications for vehicular networks has expanded hugely. Thus, the share of available network resources such as connectivity and computing power should be revisited to support the novel mobile applications in cloud-enabled vehicular networks [236]. The challenges to solve are: discover the more suitable communication link (i.e. with adequate data rate, low jitter and errors); support multi-hop communications; ensure data privacy; give incentives to vehicles to cooperate and either relay traffic from others or cache data for other vehicles; support distinct patterns of mobility; use the parked cars as serving nodes (i.e. like BSs or APs) for the cars circulating on the road; and increase the comfort of the passengers of autonomous vehicles, giving them all the Internet contents they aim for (e.g. streaming movie, live concert). Finally, new solutions are needed to handle the large amounts of data in vehicular networks [237].

The following two vehicular scenarios show that the usage of GT can bring positive outcomes. In the first one, the management of both radio and computational resources are pertinent. Envisioning these aspects, we propose a CG augmented with RL to minimize both task processing cost and the latency. In the second scenario, mobility support for vehicles without disrupting any of the diverse services being used by each vehicle is very challenging. We envision proactive data distributed caching along the expected route each vehicle travels towards a well-known geo-located destination. The services running inside the car can have access to this data, minimizing the access delay and disruption. We propose a hierarchical model formed by several layers. At the bottom, we have intelligent poles located along a single street of a city. These poles play a NC game with RL, which scales well in dynamic systems with randomness. The street poles are also the followers of a SG, where the corresponding master is a MEC device responsible for a city block (set of nearby streets). Overlaying the previous SG there could be another SG model, where the followers of the latter model are the masters of some SGs operating at the bottom level. These masters are each responsible for a set of city blocks. In theory, this hierarchical architecture could offer to the entire city a proactive and scalable service of distributed data caching. The number of layers in the system is limited by the maximum convergence time to the wished stable state.

Unmanned autonomous vehicles

Some examples of unmanned vehicles are aerial [159][238][65] and maritime [239][240]. UAVs were initially developed for military monitoring and surveillance tasks but found several interesting applications in the civilian domain. A promising application/technology is to use drone small cells (DSCs) to expand wireless communication coverage on demand [159]. This requires wireless communication links with sufficient QoS values (e.g. in throughput, delay, jitter), satisfying security requisites (e.g. avoid fake DSCs), and optimizing the energy consumption [165]. An interesting idea for UAVs is to deploy a crowd surveillance system [238]. Another tantalizing perspective is using unmanned vehicles for maritime tasks [239][240] such as surveillance and patrolling, aquaculture inspection, or wildlife monitoring. The main goal is set up a system where a vehicle cooperatively engages with others in a way that all fulfil a specific common mission. Organizing vehicles in clusters enables cooperative learning in each cluster and supports system scalability.

We now outline two scenarios using UAVs where GT can bring positive outcomes. In the first of these, we note that mobile operators are deploying ultra-dense cells to alleviate the lack of cellular resources. Nevertheless, due to some temporary network physical impairments, e.g. available channels and/or receivers' interference, the deployment strategy may not always be sufficient to grant a communications service with enough quality. In this case, we propose a mesh network formed by UAVs that fly over either a large city at busy times or crowded areas, e.g. live musical concerts, stadium sports events. As an example, in busy times, the most crowded areas of a city cause the associated parts of the terrestrial mobile network to be terribly congested. Some mobile traffic can be offloaded from those congested areas to the overlay wireless mesh network formed by flying UAVs. Then, the offloaded traffic can be routed towards a flying gateway Femtocell that injects it again in a less busy part of the ground mobile network. To model this, we propose a NC game with RL played among the diverse UAVs. The second scenario is about replacing, as fast as possible, a terrestrial network infrastructure that has been destroyed by e.g. an earthquake. A UAV wireless mesh could be the unique network available for the rescue team. Consequently, the mesh resources need to be used very efficiently. We propose a Stackelberg Game, where the leader is the gateway to the Internet and the followers are the flying UAVS of the mesh.

Other Scenarios

Recently, electric vehicles (EVs) and plug-in hybrid EVs have been considered as natural components of future electricity power systems, due to their efficient integration, cost savings, and environmental advantages. This is challenging the design of power systems in: scheduling energy charging plans for connected EVs, guaranteeing energy demands of EVs during peak hours, and exchanging information among EVs and the grid (or aggregators). Another interesting area to investigate is intrusion detection/prevention and mitigation of DDoS attacks. Recent literature about micro-grids is available in [241][242][243][244][245][246].

The Tactile Internet (TI) involves human-to-machine (H2M) communications for remote supervision of tactile/haptic devices. These devices have local sensors and actuators, and are managed via the Internet in a completely reliable and deterministic way [247]. The TI seems to converge towards a well-defined list of design goals [248]: very low latency in the order of 1ms; ultra-high (several nines) reliability; H2H/M2M coexistence; and high security. Other important challenges for the TI are [248][249]: resource management; task allocation and orchestration; mobility of robots; remote robot steering and control applications. In [80] the authors studied a solution to accommodate, within the same network, latency-sensitive TI and bandwidth-intensive traffic types.

## 4 Research Directions for Theoretical Model Games and Multi-Access Edge Computing

We have identified some interesting future research directions for GT related to the MEC use cases discussed earlier. These cover augmented reality, cognitive assistance, dynamic and fair (virtual) resource allocation, seamless session transfer, and congestion control. The aspects of augmented reality and cognitive assistance require location and environmental information obtained via camera/sensors. Dynamic and fair resource allocation makes a heavy demand on elastic system capabilities such as storage, available energy, communications capacity, traffic steering, computing, and intelligent agents (IAs). An IA is an autonomous entity which is formed by two internal elements, performance and learning. The performance element of an IA obtains raw data from sensors; it applies a set of policies (i.e. if-then rules) to the acquired data, creating a refined perception of the environment; it controls the environment via actuators; and it "waits" for the next raw data from sensors to initiate a new processing cycle. The learning element of an IA determines how the policies, which are being used in parallel by the performance element, should be modified to the IA perform better in the future. The seamless session transfer requires efficient support of ubiquitous services as well as a suitable life cycle of virtualized resources. Control congestion is a mandatory mechanism to avoid eventual network breakdowns.

The relevant technical networking aspects that mitigate congestion and empower the performance of MEC systems are traffic offloading [250][11][251], D2D communication [170][54][9], relaying [252][196][134], novel pricing schemes for peer-to-peer sharing platforms under incomplete information of users' service valuation [253], and Full Duplex (FD) communications [254]. The latter could enhance the performance of upcoming 5G wireless access technology [200][255][256]. In fact, a FD system allows a specific node to send and receive the transmitted signals at the same time and using the same frequency. This doubles the spectral efficiency on each wireless link. The authors of [254] have used both GT and matching theory to analyse diverse centralized and distributed FD communication networks. Another way to avoid congestion is to deploy mechanisms of cooperation among the devices. As already discussed, CGs can enhance the cooperation among devices [31][183][171]; but CGs of course have a cost [186]. The cost is associated with the extra signalling traffic to manage coalitions. This extra traffic depletes terminal battery energy, adds additional delay due to the information exchange, and could introduce security vulnerabilities. These are open research issues for GT. In addition, trust and authentication require more investigation, essentially for distributed and scalable solutions. The presence of malicious (or colluding) entities and their possible attacks against incentive mechanisms for cooperation need further research. GT is useful for detecting and counteracting these attacks [257]. In [258] the authors propose a real-time dynamic congestion-sensitive tariff that tracks trends of load or interference, maximizing network efficiency and fairness.

In terms of the deployment of theoretical games at the network edge, the use of a single game to control the operation of such a heterogeneous and dynamic data communication system is difficult, because it displays dynamic behaviour, constrained resources, uncertain information, as well as several simultaneous critical trade-offs to balance, namely: available energy vs. relaying traffic, computation offloading vs. channel interference, and data consistency vs. latency. So, we envision an alternative way to control the system. To support this, at the MEC servers, we run several virtual machines or "lighter" namespaces (e.g. Docker, Linux containers), which Fig. 11 designates as Edge Servers (ESs). Each ES is formed by a SDN [16] controller with a Southbound API that controls the network devices within the domain associated with that controller. In parallel, the SDN controller has a Northbound API that enables top-level agents (or NFVs [259]) to orchestrate the network resources. These agents run theoretical games with specific goals, acting like decision-makers. The agents' decisions could also enter in conflict, deprecating network efficiency. To overcome this problem, service function chaining [260] prioritizes and coordinates the decisions among all the agents, enabling a stable orchestration of resources. This orchestration is the fully-integrated distributed control of network, computational, and storage resources at the network edge. This can be designated as Software Defined Mobile Edge Computing (SD-MEC). In addition, SDN controllers can combine header fields from any stack layer, creating a cross-layer design that is suitable for emerging wireless scenarios such as D2D [261], vehicular [262], and sensor networks [263] [264]. This system can have a hierarchical design to support reliable, scalable, and energy-efficient power control for D2D communications [265]. The cloud broker units in Fig. 11 are Network Address Translation middleboxes [266]. Fig. 11 also shows several communities; each one could be formed by people, devices, or systems. These entities are within communities because every entity has strong relationships with others in one form or another. This is the basis for another emerging concept on networking designated as 'mobile social networks' [9][267]. A mobile social network is a user-centric system in which the devices not only process data, but also deal with context information (e.g. location) to either react (passively) or predict (actively) to respectively the network status at each time or an expected network status.

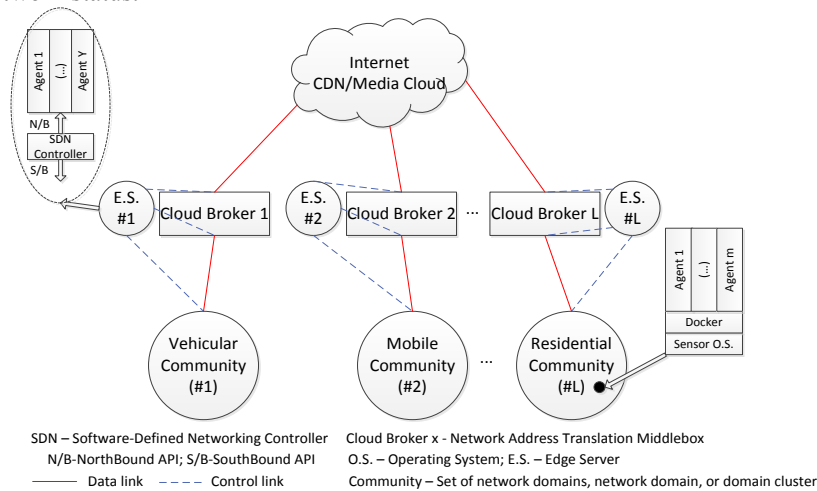

Fig. 11. Design of Mobile Social Network controlled and orchestrated by SD-MEC

Each community device (e.g. sensor) could also be running diverse agents (e.g. the already discussed IA). Each agent is a virtualized container (Docker) and is associated with a specific edge provider. The presence of agents at the network periphery can transform information into network knowledge, creating the cognitive network. This cognitive network, using theoretical games fuelled by evolutionary algorithms and artificial intelligence, can infer valid and precise human behavioural patterns from the aggregated data previously acquired via local sensors. Using the inferred knowledge, the agents – after running an election among essentially the neighbouring agents that sorts out a ranked list of viable automatic actions – select the top-most ranked actions, run these over the monitored systems, including network resources and services, and learn from the results [268]. Thus, a digital shared intelligence is created at the network periphery to successfully solve the initial paradox of offering the most highly viable service quality with minimum cost. Another trade-off to be solved is to support fairness among the competitive heterogeneous flows in the presence of high and unexpected fluctuations on the traffic load, alongside other network unpredictable trends. The authors of [269] propose a three-layer hierarchical system with sensors, gateways, and controller that enhances context measurement, processing, and communications. This hierarchical framework satisfies both traffic load and a longer lifetime for the whole system.

We emphasize that these changes at the edge domains would introduce functional changes in the way the network resources are discovered and used: i) from IPv4 to IPv6; ii) from World Wide Web (WWW)/client-server/end-to-end to Named Data Networking (NDN) [270][271]/publish-subscribe/P2P or multicast communication; iii) from network layer forwarding data by interface IP address to forward directly on a name associated to a chunk of data. These changes would occur in a gradual way from the edge towards the network core, potentially blurred when reaching a part of the network still operating in the legacy mode. In this case, Network Address Translation (NAT) middleboxes [266] (or other gateways/rendezvous servers) would be needed for interconnection purposes.

## 5 Summary and Future Trends

This paper initially provided background information, including GT and MEC, for non-specialist readers. Then we substantially reviewed game theoretical contributions for wireless data communications, focusing also on scenarios aligned with the emerging MEC model.

Our conclusion is that the GT models most suitable for use in future work to fulfil MEC research directions are coalitional and evolutionary ones. The main advantage of coalitional models is the high scalability of hierarchical proposals based on clusters. Additionally, they can bring huge savings in terms of energy consumption, essentially for remote battery-operated devices (e.g. sensors). The main advantage associated with evolutionary alternatives is the easy adaptation to unexpected network behaviors using some learned successful strategies from the past. Bayesian games also seem interesting solutions for scenarios with limited or highly variable context information. NC models are not efficient solutions for MEC future services. This can be justified due to the absence of learning and the selfishness of the players. Nevertheless, the NC model augmented, e.g. with an extra mechanism to incentivize cooperation [136], can also be a viable option for MEC services. Alternative ways to mitigate selfishness among NC players are learning [134][44][168][169][69][170][159] and machine learning [63]. An example where learning is deployed in GT models is Reinforcement Learning, which can address the typical challenges of wireless networks: viz. resource scarcity, distributed nature, uncertainty, and randomness [60]. Ref. [272] proposes a NC learning model that improves QoE for DASH/HTTP video streaming, preserving fairness among the users. The NC game can also be played among the followers of a SG. The leader of this game can run on a SDN [16] controller or a NFV instance [259].

We have highlighted many MEC scenarios. There are excellent research opportunities for applying GT to a broad range of MEC services such as: low data/service access latency; distributed offloading computing over heterogeneous and hierarchical MEC environments; proactive data caching at the network edge managed by data popularity, social user connections, and available node battery energy; low-power wireless communications and networking for IoT; fingerprinting localization technology [273][274] to support indoor location based services; high mobility wireless communications [275], such as a cloud-Radio Access Network architecture for cloud high speed train communications, where a virtualized single cell design diminishes handover failures and offers a reliable communication services [276]; and understanding emerging security and privacy problems in cyberspace and potential solutions [277].

The authors of [278] state that GT is now central to many research investigations. They argue that learning [60][63] is a promising route for improving GT's predictive power, hopefully in polynomial time. They also emphasize facets where GT becomes more useful: the game results should not only be valid but precise; the GT models should have limited complexity, and they should be easily reproduced and tested by the research community [131].

We believe that the next few years will definitively clarify what real impact theoretical model games will have on the enhancement of MEC services at the network edge in coordination with remote clouds.

## Acknowledgement

The authors are grateful to the anonymous reviewers for their constructive comments to improve the quality of the paper.